\documentclass{emulateapj}
\usepackage[greek,english]{babel}
\usepackage[latin9]{inputenc}
\usepackage{amssymb}
\usepackage{textcomp}
\usepackage{natbib}
\usepackage{rotfloat}
\usepackage[T1]{fontenc}
\usepackage{graphicx}
\usepackage{esint}
\usepackage{verbatim}

\slugcomment{Draft version January 17, 2011}

\shorttitle{Gas and Dust Toward DG Tau B and VV CrA}
\shortauthors{Kruger et al.}

\begin{document}


\title{Gas and Dust Toward DG Tau B and VV CrA}

\author{Andrew J. Kruger, Matthew J. Richter}
\affil{Department of Physics, University of California at Davis, CA 95616, USA}

\author{John S. Carr}
\affil{Remote Sensing Division, Naval Research Laboratory, Code 7210, Washington, DC 20375, USA}

\author{Joan R. Najita, Greg W. Doppmann}
\affil{National Optical Astronomy Observatory, Tucson, AZ 85719, USA }

\and
\author{Andreas Seifahrt}
\affil{Department of Physics, University of California at Davis, CA 95616, USA}

\begin{abstract}
We present findings for DG Tau B and VV CrA, two of the objects observed
in our Spitzer IRS project to search for molecular absorption in edge-on
disks, along with near-IR spectroscopy of the CO fundamental transitions
and mid-IR imaging. While the only gas absorption seen in the Spitzer
IRS spectrum toward DG Tau B is CO$_{2}$, we use gas abundances
and gas/ice ratios to argue that we are probing regions of the disk
that have low organic molecule abundances. This implies the rarity
of detecting molecular absorption toward even edge-on disks with Spitzer
IRS is a result of high dependence on the line of sight. We also argue
the disk around DG Tau B shows high amounts of grain growth and settling.
For VV CrA, we use the silicate absorption feature to estimate a dust
extinction, and model the disk with a spectral energy distribution fitting tool to give evidence in support of the disk geometry presented
by Smith et al. (2009) where the Primary disk is the main source of
extinction toward the infrared companion.
\end{abstract}



\section{Introduction}

The study of disks of dust and gas surrounding young stars provides
insight into the environment in which planets are likely to be forming.
Molecular abundances in the disk can be used to understand the chemical
evolution of the gas, disk processes such as radial and vertical mixing,
and the effects of stellar irradiation on the disk chemistry (Bergin
et al. 2007). While models predict disk chemistry in the earth-like
planet forming regions (i.e. Markwick et al. 2002; Ag\'{u}ndez et al.
2008; Glassgold et al. 2009; Woods \& Willacy 2009), supporting observations
of the chemical tracers have been scarce to date. 

The Spitzer Infrared Spectrograph (IRS) made observations that probed
the chemistry of the inner disk, detecting molecular absorption (Lahuis
et al. 2006) and emission (Carr \& Najita 2008; Salyk et al. 2008).
The resolution of IRS was well-suited for detecting molecules with
large equivalent widths from ro-vibrational Q-branches. Lahuis et
al. (2006) detected warm C$_{2}$H$_{2}$, HCN, and CO$_{2}$ in absorption
toward IRS 46, likely originating in the inner regions of a circumstellar
disk seen near edge-on. Unfortunately, systems displaying molecular
absorption are rare with, for example, IRS 46 being the only one of
roughly 100 young stellar objects (YSO) observed in the c2d Spitzer
Legacy program (Evans et al. 2003). 

Molecular C$_{2}$H$_{2}$ and HCN absorption have also been detected
toward GV Tau N (Doppmann et al. 2008; cf. Gibb et al. 2007), a classical
infrared companion (IRC; Koresko et al. 1997). IRCs are binary companions
to young stars that are faint in the visible but dominate the system
flux in the infrared. Correia et al. (2007) found with MIDI/VLTI data
that three IRCs, including GV Tau, are likely T Tauri stars (TTS)
with the disk seen almost edge-on. Doppmann et al. (2008) found the
warm temperature and the relative velocity shifts of the molecular
absorption further supported the edge-on disk model. Gibb et al. (2007)
compared the molecular absorption abundances to predictions made by
the chemical disk model of Markwick et al. (2002) and found them to
be consistent with the inner region of a protoplanetary disk.

Because IRS 46 and GV Tau N show the rare features that are valuable
for probing the inner disk chemistry, it is advantageous to study
more disks with similar attributes. We were granted time with Spitzer
IRS to obtain high signal-to-noise (S/N) spectra of seven YSOs. Our
target selection was limited to systems thought to be similar to IRS
46 and GV Tau N with a large column of gas between the observer and
the emitting source necessary for absorption studies: three circumstellar
disks seen near edge-on and four classical IRCs. DG Tau B and VV CrA
are two of the objects observed in this project.

Mundt \& Fried (1983) first designated DG Tau B as the source of a
Herbig-Haro jet (HH 159) that flows at 15$^{\circ}$ to the plane
of the sky. Near-infrared HST images show DG Tau B to be a bipolar
reflection nebula embedded in an envelope (Stapelfeldt et al. 1997).
Stark et al. (2006) modeled near-infrared disk images, finding it
likely has a disk inclination of $\sim$73$^{\circ}$. Previous Spitzer
IRS measurements have shown this system displays CO$_{2}$ ice (Watson
et al. 2004) and gas absorption (Pontoppidan et al. 2008), so we obtained
higher S/N to search for other species.

The northeast component of the VV CrA system is a classical IRC, and
Correia et al. (2007) used MIDI/VLTI observations to show the IRC
is likely a TTS seen near edge-on. Smith et al. (2009; hereafter S09)
used CRIRES/VLT to detect CO gas absorption with multiple temperature
components toward the Primary and IRC. They proposed two disk morphology
scenarios to explain the CO absorption seen toward VV CrA IRC: (A)
the gas absorption is in a disk around the Primary in the line of
sight to the IRC or (B) all the CO absorption is in a disk around
the IRC. While S09 considers Case A to be more likely, with the observed
CO residing in the warm surface layer and cooler midplane of the Primary
disk, neither case could be dismissed so we consider both geometries.

In addition to the IRS spectra, we observed both targets with ground-based
telescopes. We used M-band spectroscopy to observe the CO fundamental
($\Delta$v=1) transitions toward both objects. Observing these lines
probes a wide range of radii in circumstellar disks and has proven
useful for exploring the disk gas in the inner disk regions (Najita
et al. 2003; Brittain et al. 2007 and references therein). Finally,
because molecular absorption is limited to studies in the line of
sight to the continuum emitting source and the Spitzer IRS spectrum
combines both VV CrA components, we obtained mid-IR images of VV CrA
in order to map the continuum at spatial scales unavailable with Spitzer
and to separate the continuum flux for the two components.

We present here results for DG Tau B and VV CrA from our study of
sources in the Spitzer survey. We include high-resolution absorption
spectroscopy and, for VV CrA, mid-IR imaging. In $\S$2 we introduce the
observations taken; in $\S$3 we describe the data reduction and discuss
the molecular detections; and we discuss modeling and the conclusions
in $\S$4.

\section{Observations}

\subsection{Spitzer IRS}

We observed DG Tau B and VV CrA as part of our Cycle 5 General Observers
project, Program ID 50152. We used the Spitzer Infrared Spectrograph
(IRS; Houck et al. 2004) short-low and short-high modules (SL and
SH) in stare mode to search for molecular absorption species with
large equivalent widths such as ro-vibrational Q-branches. The SL
spectra obtained have a coverage 5.2-14.5 \textgreek{m}m with R=$\lambda/\Delta\lambda$=60-127,
and the SH spectra have a coverage of 9.9-19.6 \textgreek{m}m with
R=600. We adopted the observing technique described in Carr \& Najita
(2008). In order to verify the repeatability of the observations,
we lowered the exposure time for each data frame and increased the
number of integrations. Each setting (SL1, SL2, SH) thus had twelve
on-source data frames at each nod position, each with a 6 second integration.
For hot pixel subtraction and to remove background emission, each
observation was supplemented by off-source integrations for half the
time as on-source (Carr \& Najita 2008). DG Tau B (AOR: 25680128)
was observed on 2008 November 5 at 4$^{h}$27$^{m}$2.66$^{s}$ +26$^{\circ}$5$^{\prime}$30.5$^{\prime\prime}$
(J2000). VV CrA (AOR: 25680896) was observed on 2008 November 18 with
the pointing on the IRC at 19$^{h}$3$^{m}$6.86$^{s}$ -37$^{\circ}$12$^{\prime}$48.2$^{\prime\prime}$
(J2000). The coordinates for VV CrA were chosen to center the slit
on the IRC, and were calculated using a 2.1$^{\prime\prime}$ offset,
44$^{\circ}$ east of north (Chelli et al. 1995) from the known primary
coordinates found on SIMBAD. Both observations used the nominal pointing
accuracy of 0.1$^{\prime\prime}$.

\subsection{Infrared Spectroscopy}

On 2008 November 16, we used NIRSPEC (McLean et al. 1998) on Keck
II in echelle mode with R=25,000 to detect the CO fundamental transitions
toward DG Tau B. Our instrument setup included the M-wide blocking
filter, along with echelle and cross disperser settings to image orders
14-16 so we could view the $^{12}$CO v=0-1 R(0)-R(9) and P(25)-P(30)
transitions. We used the standard ABBA nod sequence with thirty 1-second
exposures co-added at each position. Standard star observations were
taken before and after DG Tau B for telluric division.

For a similar observation of both VV CrA components, we used Phoenix
(Hinkle et al. 1998) on Gemini South on 2009 April 6. We used the
standard four pixel slit (0.35$^{\prime\prime}$), giving R=50,000.
The M2150 filter and instrument settings were used to detect the $^{12}$CO
v=1-0 R(3)-R(5) transitions near 4.66 \textgreek{m}m. Each ABBA nod
position had a 120 second integration (four 30 second exposures co-added
internally), and we used a telescope position angle of 44$^{\circ}$
to place both binary components within the slit. Standard star observations
were taken after the science frames for telluric division.

\subsection{Imaging}

We imaged the binary components using T-ReCS (Telesco et al. 1998)
on Gemini South using 6 filters ranging from 7.7-25.5 \textgreek{m}m
(see Table 1). We cycled through each of the filters while staying
on the science target, and took calibration images of standard stars
before and after the science target for point spread function (PSF)
corrections and to test for any changes in PSF or calibration during
the target observation. Observations were conducted using the standard
chop-nod mode. When we nodded, the target was placed at the center
of the field of view for half the total integration time, and the
other half placed the target into a corner of the array for frames
to be used for background subtraction.

\section{Data Reduction and Detections}

\subsection{Spitzer IRS}

We used the reduction technique described in Carr \& Najita (2008).
We used the integrations of the nearby blank sky to identify hot pixels,
which were removed from the science frames by interpolation using
custom routines in IDL. We then used IRAF to extract the raw spectra
from each individual raw data frame. A similar procedure was used
to extract spectra from the regularly observed HR 6688, which we used
for standard star division, to remove fringing, and for flux calibration.
Because the fringing in each observation is dependent on the target
position on the slit, each data frame was divided by the standard
star observation that showed the least amount of fringing in the shorter
wavelengths after division. Any residual fringing was removed using
IRSFRINGE in IDL. Calibration of the SL spectra is highly dependent
on the slit position, so we used the optimal point source extraction
in the Spitzer IRS Custom Extractor (SPICE; version 2.2) to extract the individual
spectra before co-adding. 

The Spitzer IRS spectrum of DG Tau B, with an average S/N=100 per
pixel, can be seen in Fig. 1. A 10\% continuum flux difference between
the SH and SL modules was corrected by scaling the SH by a constant,
without any adjustment to the continuum shape, to match the SL at
wavelengths where the orders overlap. The spectrum shows the amorphous
silicates, CO$_{2}$ ice absorption, water ice, and an absorption
feature at 6.8 \textgreek{m}m, which is possibly due to methanol (Watson
et al. 2004, Zasowski et al. 2009), as well as the CO$_{2}$ gas absorption
at 15.2 \textgreek{m}m as previously recognized by Pontoppidan et
al. (2008). We further detect {[}NeII{]} emission at 12.81 \textgreek{m}m
with line intensity $3.0\pm1.5\times10^{-15}\mbox{ ergs cm}^{-2}\mbox{ s}^{-1}$.

The Spitzer IRS spectrum of VV CrA shown in Fig. 1, with S/N=160 per
pixel, is of the combined binary components as they were unresolved
within the slit. It can be seen there is a continuum mismatch between
SH and SL modules, likely due to being centered on the IRC and from
the modules having different slit position angles and widths. The
SH and SL modules have position angles 221.5$^{\circ}$ and 136.7$^{\circ}$
(for a Spitzer roll angle of 0$^{\circ}$), respectively, so during
the SH observation the binary components were separated 0.31$^{\prime\prime}$
across the 4.7$^{\prime\prime}$ slit while during the SL observation
the components were separated 1.20$^{\prime\prime}$ across the 3.6$^{\prime\prime}$
slit. The SL1 is at an intermediate wavelength and flux, so we scaled
the SL2 and SH modules by a constant to match the intermediate SL1
continuum. We find the system displays amorphous silicate absorption
and CO$_{2}$ ice absorption at 15.2 \textgreek{m}m.

\subsection{CO Fundamental}

To reduce our Phoenix and NIRSPEC spectra, we used standard spectral
reduction routines in IRAF. After flatfielding each frame, we subtracted
nod pairs to remove the sky emission. The target spectra were extracted
using \emph{apall} in IRAF with background subtraction using both
sides of the profile (from both sides of the total binary profile
for VV CrA). In order to increase S/N and for better telluric division,
we first extracted the spectrum from each individual science and standard
star frame (standards listed in Table 1). We then divided each science
spectrum by each standard star spectrum using their corresponding
airmass, then co-added the spectra. We found that applying telluric
division before co-adding the spectra resulted in better telluric
correction. The wavelength solution was made by using the IRAF task
\emph{identify} on sky emission lines in coadded science frames. The
rest frame wavelengths of the emission lines were taken from the HITRAN
database (Rothman et al. 1998).

Our NIRSPEC spectra of DG Tau B, seen in Fig. 2, have S/N=21 and 28
per pixel for orders 15 and 16, respectively, and we found both $^{12}$CO
and $^{13}$CO in absorption (see Table 2). The average line profiles
for the low- and high-J $^{12}$CO transitions have slightly different
velocity shifts and widths (see Table 3 and Fig. 3) which indicates
we are likely probing multiple velocity components.

We were able to clearly resolve the VV CrA components with Phoenix,
and the spectra, shown in Fig. 4, had an average S/N of 220 per pixel
for the IRC and 280 for the Primary. The VV CrA IRC displays $^{13}$CO
and optically thick $^{12}$CO in absorption (see Table 4), as well
as a shallow, broad $^{12}$CO absorption feature with $\mbox{v}{}_{LSR}=-24\mbox{ km s}^{-1}$
and $\mbox{FWHM=32.9 km s}^{-1}$ (see line profile in Fig. 5). The
VV CrA Primary shows cold $^{12}$CO within broader $^{12}$CO emission
lines. The cold, unresolved $^{12}$CO absorption lines have $\mbox{FWHM=5.35 km s}^{-1}$
while using the 4-pixel slitwidth with resolution $\mbox{R=6 km s}^{-1}$,
suggesting our resolution was limited by the PSF of the target rather
than the entire slit width.

The Primary and IRC both also show a complicated emission structure.
This structure can be understood as a blend of broad CO emission lines
from numerous ro-vibrational levels. Both of the VV CrA components
show CO overtone emission (Prato et al. 2003) with emission up to
at least the v=4-2 vibrational band. The observed CO fundamental bands
are consistent with the expectation of a hot, optically thick emission
spectrum. In this respect, the spectra are similar to the extreme
T Tauri star RW Aur (Najita et al. 2003).

\subsection{Imaging}

The T-ReCS images show a resolved VV CrA binary (Fig. 6). Images were
made by subtracting nod pairs and co-adding the resulting images.
High winds caused poor image quality, so we created custom IDL routines
to center each data frame on the peak of the primary companion before
co-adding. This gave better final image quality, but we did not detect
any extended emission. We measured photometric fluxes using a 2.1$^{\prime\prime}$
aperture centered on each component and on the standard star observations,
which were then used for flux calibration (see results in Table 5).
While the primary does show a hint of silicate absorption, a majority
of this feature seen in the Spitzer IRS spectrum originates in the
IRC. We find that the broadband fluxes measured with T-ReCS fit the
Spitzer IRS spectrum if we assume we are collecting only 70\% of the
flux from the Primary (see Fig. 1), with the flux loss likely due
to slit losses from being centered on the IRC.

\section{Discussion}

\subsection{Molecular Detections}

\subsubsection{DG Tau B}

All of our models of the CO and CO$_{2}$ gas were done using a single
temperature slab model. We fit the $^{13}$CO R(16-19) transitions
with the assumption the transitions are optically thin and found a
temperature of 288\textpm{}7 K with column density $\mbox{N(}^{13}\mbox{CO)=}3.5\pm0.6\times10^{17}\mbox{ cm}^{-2}$
(see population diagram in Fig. 7). If we fit the high-J $^{12}$CO
P(25-27, 30) lines to an optically thin model as well, we find the
gas would be at a high temperature of $\mbox{930}\pm60\mbox{ K}$
with $\mbox{N(}^{12}\mbox{CO)=7.86}\pm0.2\times10^{17}\mbox{ cm}^{-2}$.
However, this would indicate a $^{12}$CO/$^{13}$CO ratio of only
2.2, much lower than the mean ratio found in the local
ISM (Wilson 1999). This indicates all of the observed $^{12}$CO lines
are likely optically thick. To constrain the temperature of the absorbing
gas, we modeled $^{12}$CO gas with a column density $2.42\times10^{19}\mbox{ cm}^{-2}$,
estimated using the N($^{13}$CO) and mean isotopic ratio of 69$\pm$6 from Wilson
(1999), and varied the turbulence and temperature. We could not
model all the $^{12}$CO transitions to a single temperature, with
the low-J transitions always showing higher optical depths than permitted
by fits to the high-J transitions. This is likely due to multiple
velocity components being viewed in the low-J transitions, which
would account for the difference in velocity shift and width from
the high-J transitions. We thus fit only the high-J P(25-27, 30) transitions,
and the goodness-of-fit, shown in Fig. 7, indicates the gas is likely
$370\pm20\mbox{ K}$ with a Doppler b-value $\mbox{v}_{turb}=2.1\pm0.2\mbox{ km s}^{-1}$.
If the $^{12}$CO is self-shielding and the $^{12}$CO/$^{13}$CO
is raised to 100, similar to the ratios found in S09, these values
would not change much as the temperature would still be $\sim350\mbox{ K}$
and $\mbox{v}_{turb}\sim2\mbox{ km s}^{-1}$.

We find the model for the CO$_{2}$ gas has degenerate fits by changing
the microturbulence and temperature, so we use the Doppler b-value
of $2.1\mbox{ km s}^{-1}$ found with the CO gas, and find $\mbox{T=300 K}$
and $\mbox{N(CO}_{2})=2.9\times10^{16}\mbox{ cm}^{-2}$. We further
find upper limits for C$_{2}$H$_{2}$ and HCN, two other organic
molecules detected toward IRS 46 (Lahuis et al. 2006), to compare
molecular abundance ratios. To get upper limits on column densities,
we note that temperatures of both the CO and CO$_{2}$ gases toward
DG Tau B are similar to that found toward IRS 46 (400 and $\mbox{300 K}$,
respectively), so we start with the assumption that if C$_{2}$H$_{2}$
and HCN are present, they would have similar temperatures to IRS 46
(700 and 400 K, respectively). To derive the upper limits, we individually
modeled the absorption spectrum for the molecules at those temperatures.
We then used the model-generated profiles to measure feature strengths
at various wavelengths in the DG Tau B spectrum,
giving us a distribution of feature strengths we would expect to measure
given only the spectral noise. We then set the upper limit
as the column density that would provide a 3$\sigma$ integrated intensity.
We find the upper limit on the C$_{2}$H$_{2}$/CO$_{2}$ ratio (< 0.19) is
not very different from the ratio measured for IRS 46 (0.3),
suggesting the non-detection of C$_{2}$H$_{2}$ in our data may be
the result of the lower absorbing column in this source. The upper
limit on the HCN/CO$_{2}$ ratio (< 0.22) is a factor of two smaller than was
measured for IRS 46 (0.5). Modeling the gas at higher temperatures with
similar column densities decreases the integrated intensity of the
absorption features, so higher gas temperatures would result in higher
upper limits on the column densities. A temperature increase of $\mbox{50 K}$
would raise the column density upper limit by $\sim$15\% for both
molecules. If we model the molecules at $\mbox{300 K}$, like the
CO$_{2}$ gas, the upper limits of the C$_{2}$H$_{2}$/CO$_{2}$
and HCN/CO$_{2}$ ratios would be 0.06 and 0.18, respectively.

The {[}NeII{]} emission, as seen toward DG Tau B, is a feature commonly
found by Spitzer IRS toward TTSs (Lahuis et al. 2007). G\"{u}del et al.
(2010) showed that among TTS the {[}NeII{]} emission strength increases
with the jet mass loss rate, the latter as inferred from the high
velocity component of the {[}OI{]} emission. Mundt \& Fried (1983) found
that HH 159, a jet associated with DG Tau B, has a mass loss rate
of $10^{-9}\mbox{ M}_{\bigodot}\mbox{ yr}^{-1}$. If we assume DG
Tau B is $\mbox{145 pc}$ away, our measured {[}NeII{]} luminosity
of $7.5\times10^{27}\mbox{ ergs s}^{-1}$ is similar to that of TTS
with comparable mass loss rates (cf. Fig. 5, G\"{u}del et al. 2010). 
As we do not see C$_{2}$H$_{2}$ and HCN in the disk, a simple explanation is they have low abundances in the disk.  
This could possibly be due to photodissociation.
The {[}NeII{]} emission may be an indication the gas is irradiated
by X-rays (Glassgold et al. 2007), and/or UV irradiation (Hollenbach
\& Gorti 2009), which in turn could reduce the C$_{2}$H$_{2}$ and
HCN in the disk.  However, many T Tauri stars with {[}NeII{]} emission
similar to or stronger than that of DG Tau B show emission
in C$_{2}$H$_{2}$ and HCN (Carr \& Najita 2011, in prep).

Alternatively, as C$_{2}$H$_{2}$ and HCN are commonly found in circumstellar
disks (Pontoppidan et al. 2010), these species could be in the disk
but simply not in our line of sight. This may be implied by the most
striking difference between DG Tau B and IRS 46: while the column density 
of CO toward DG Tau B is only 10 times that toward IRS 46, the gas ratios to CO
 in DG Tau B are 1\%-2\% of that found toward IRS 46 (see Table 6).
Also, the limits on the ratios N(C$_{2}$H$_{2}$)/N(CO) and N(HCN)/N(CO) are much smaller for DG Tau B than the measured ratios toward GV Tau N (Gibb et al. 2007).  While the large CO column toward DG Tau B implies we are probing a large amount of disk material, we are not detecting the large columns of other gaseous species.


The disk chemistry model in Ag\'{u}ndez et al. (2008) may provide further insight
into the disk conditions probed by our observations. They focus their
models on the photon dominated region of the disk, and use photochemistry
to estimate the steady state abundances of simple organic molecules.
They show the C$_{2}$H$_{2}$ and HCN have low abundances for disk
radii $\mbox{>1.5 AU}$, and the CO/CO$_{2}$ is large for radii $\mbox{<1 AU}$.  While the 
temperatures 
we measured for CO and CO$_{2}$ are cooler than expected for radii <1 AU, the
vertical structure of these molecules is not available. One possible
scenario is the line of sight passes below the main temperature inversion
layer.   

Doppmann et al. (2008) note GV Tau N has a larger $\tau_{SiO}/\tau_{CO_{2}ice}$ than other Class I YSOs in Taurus (typically $\le4$; Furlan et al. 2008), and suggest this could happen if a majority of the probed silicate is residing in a warm region of the disk atmosphere.  We calculate this ratio for DG Tau B and IRS 46 for comparison.
To do this, we used a polynomial fit across the features to determine
the initial continua.  For the silicate feature, we fit to the continuum in the 13-14.8  
\textgreek{m}m region, and to the continuum between the water and "methanol" ice at 
wavelengths <7.8 \textgreek{m}m.  For the CO$_{2}$ feature, 
we assume the CO$_{2}$ ice is located at larger disk radii than the silicate, as 
found by Watson et al. (2004), so the ice is absorbing from a continuum already 
including the silicate absorption features.  We thus use a continuum for DG Tau B that is 
modified from that used in Watson et al. (2004), found by fitting the spectrum at wavelengths 
13-14.8 \textgreek{m}m and >15.6 \textgreek{m}m.  The continuum fits are shown in Fig. 1.  We then used the relation $\mbox{I}_{obs}=\mbox{I}_{em}\mbox{exp}(-\tau)$,
where I$_{obs}$ is the observed flux and I$_{em}$ is the emitted
continuum, to find $\tau_{SiO}=1.52\pm0.06$ and $\tau_{CO_{2}}=0.24\pm0.01$.
Lahuis et al. (2006) found IRS 46 to have $\tau_{SiO}=0.86$, and
Pontoppidan et al. (2008) shows $\tau_{CO_{2}}=0.1$.  While both IRS 46 and DG Tau B have higher ratios than found for Class I objects in Taurus, both GV Tau N and IRS 46 have higher ratios ($\sim9$ and 8.6, respectively) than DG Tau B (6.6$\pm$0.4). Additionally, using
the CO$_{2}$ ice column density measurements for IRS 46 and DG Tau B found
in Pontoppidan et al. (2008), we find IRS 46 has a higher CO$_{2}$
gas/ice ratio, about 8 times that of DG Tau B.  This suggests the line of sight toward IRS 46 is predominantly probing warmer regions than toward DG Tau B, resulting in the detection of C$_{2}$H$_{2}$ and HCN.

Alternatively (or in addition), the ratios of CO$_2$ gas, ice and silicate could simply be a factor of disk inclination.  If the lines of sight for IRS 46 and GV Tau N pass higher in the disk atmosphere than for DG Tau B, raising the silicate to ice ratio, it could also be probing a moderately different region of the disk where the organic molecules reside.  If we are only viewing the cold to moderate temperature ($<400\mbox{ K}$) regions deeper in the disk in DG Tau B, the C$_{2}$H$_{2}$ and HCN production would be lower in these regions, thus lowering the steady state abundances (Ag\'{u}ndez et al. 2008) and resulting in their non-detection. If this is the case, the detection of pre-biotic molecules in absorption may be highly sensitive to the line of sight, even for edge-on disks.  This could possibly further explain the rarity of molecular absorption among Spitzer IRS spectra of young stars.

\subsubsection{VV CrA}

Our Phoenix spectra of the VV CrA system shows $^{12}$CO transitions
in absorption toward both components. However, the optically thick
lines for the IRC and few lines for the primary make it difficult
to extract accurate temperature and column density measurements. We
therefore use the CO abundances reported in S09 for the IRC. To get
the temperature and column density of the cold $^{12}$CO toward
the primary, we obtained the CRIRES spectra used in S09 from the ESO
archive. We reduced the data using a similar procedure to that used
for the Phoenix and NIRSPEC, although the wavelength calibration and
telluric division were performed using a telluric model fitter (Seifahrt
et al. 2010). We find the cold gas toward the primary has a column
density of $2.6\pm0.4\times10^{15}\mbox{ cm}^{-2}$ and temperature
$29\pm4\mbox{ K}$ (population diagram is shown in Fig. 8). While
doppler velocity shifts of the $^{12}$CO v=(2-0) absorption lines
reported for the IRC in S09 are $\sim3-5\mbox{ km s}^{-1}$, the CRIRES
spectrum shows a smaller shift for the v=(1-0) lines ($\sim1\mbox{ km s}^{-1}$),
similar to the shift we found with the Phoenix spectrum ($0.78\pm0.5\mbox{ km s}^{-1}$).
One change between the Phoenix and CRIRES spectra is the CRIRES spectrum
does not show the broad, shallow lines, but rather displays other
$^{12}$CO lines in the -40 to -$80\mbox{ km s}^{-1}$ range (see
Fig. 9). Thus, these broad lines are in a turbulent, varying region,
possibly located in an outflow.

While the near-IR spectrum of VV CrA is quite complicated, the Spitzer
IRS spectrum of the VV CrA binary shows only amorphous silicate and
CO$_{2}$ ice absorption. To determine the optical depths of the silicate
and CO$_{2}$ ice, we used the broadband fluxes found with the T-ReCS images
to estimate continua for the two components. The T-ReCS images indicate
almost all of the silicate absorption is toward the IRC, so this
feature is not associated with the ISM. To get the optical depth of
the silicate and CO$_{2}$ ice, we first subtracted the Primary continuum
from the Spitzer IRS spectrum, then we used the same technique used
to get the optical depths for DG Tau B. The peak optical depth for
the 9.7 \textgreek{m}m feature seen toward the IRC is $\tau_{SiO}=1.35\pm0.04$.
Using the relation $\mbox{A}_{V}/\tau_{SiO}=16.6\pm2.1$ from Rieke
\& Lebofsky (1985) for grains in the ISM, we estimate the extinction
to be $\mbox{A}_{V}=22.4\pm3\mbox{ mag}$, similar to the estimate
by Koresko et al. (1997) of $\mbox{19 mag}$.

For the CO$_{2}$ ice feature, if all of the absorption for this feature
is associated with the IRC, the optical depth would be $\tau_{CO_{2}}=0.055\pm0.004$,
which would give a ratio $\tau_{SiO}/\tau_{CO_{2}}=24.8\pm2$. 
At first glance, this high ratio could imply an edge-on disk geometry when comparing to the Class I YSOs in Taurus (Furlan et al. 2008).
However,
the $\tau_{SiO}/\tau_{CO_{2}}$ ratio found for the IRC is typical
of values found by Gibb et al. (2004) for embedded, massive YSOs,
and found by Alexander et al. (2003) for low- and intermediate-mass
YSOs. Further, the selection in Alexandar et al. (2003) included YSOs
in the CrA cloud, so an edge-on geometry may not be necessary for
the high $\tau_{SiO}/\tau_{CO_{2}}$ ratio seen toward the IRC.

\subsection{SED Model Fits}

We want to characterize the disk structures of DG Tau B and VV CrA
in order to have a context for the gas absorption, so we used the
online SED fitting tool described in Robitaille et al. (2006) to get
estimates on the disk structure. This tool finds a best fit to an
input spectral energy distribution (SED) using a database of 200,000
pre-computed radiation transfer models of circumstellar disks, with
14 physical parameters varied at 10 different disk viewing angles.
This modeler also estimates the interstellar and circumstellar extinction
separately. The user is able to input spectral points, along with
the apertures used in the measurements, at wavelengths ranging from
the UV to radio. The user can also use upper and lower limits with
a \textquotedblleft{}confidence\textquotedblright{} percentage that
rates how much to penalize models that exceed the limit. The output
disk models can be limited to those within a distance and interstellar
extinction range. The models are returned in order of increasing $\chi^{2}$
value, and while a $\chi^{2}$ cut-off is arbitrary and impossible
to make formal when fitting well behaved, symmetric models to less
behaved YSOs (Robitaille et al. 2007), we find for our disks the SED
modeler provides good fits up to $\chi_{best}^{2}-\chi^{2}\le10\mbox{N}$.

While Stark et al. (2006) created synthetic images to model the envelope
and disk structure seen in HST near-IR images, giving them the advantage
of directly modeling images that show disk structure, we can compare
and contrast to the model estimates from the SED fitter. This can
provide insight into the model results for VV CrA. If we compare the
model extinction for VV CrA to our extinction estimates using the
optical depth $\tau_{SiO}$, we may be able to constrain the location
of the dust, whether in the disk or in the interstellar medium, and
test the two disk scenarios presented in S09.

\subsubsection{DG Tau B}

We constrain the models for DG Tau B to be between 120 and $\mbox{170 pc}$
away (Gottlieb \& Upson 1969; Elias 1978; Kenyon et al. 1994) with
interstellar extinction up to $\mbox{A}_{V}=15\mbox{ mag}$ (Whittet
et al. 2001). Our Spitzer IRS spectrum was included by inputting the
flux at every 1 \textgreek{m}m with a 10\% error. The SED model fits
can be seen in Fig. 10, and the parameter results for the best fit
model are listed in Table 7.

The models predict a disk that is not seen edge-on ($18^{\circ}$), which is not surprising
as Robitaille et al. (2007) found this SED fitting tool does not determine
the disk inclination well for most models. Robitaille et al. (2007)
used the SED fitting tool to investigate DG Tau B and found an edge-on
orientation (inclination=$87^{\circ}$) for those models with no envelope
accretion. However, when envelope accretion - as has been found for
DG Tau B by Stark et al. (2006) - was added, all inclinations were
allowed. We thus exclude disk inclinations from the reported disk parameters.

Stark et al. (2006) modeled DG Tau B with a central star with $\mbox{M}_{\star}=0.5\mbox{ M}_{\bigodot}$,
$\mbox{T}_{\star}=3800\mbox{ K}$, and minimum disk radius $0.03\mbox{ AU}$.
They calculated an envelope infall rate of $5.0\times10^{-6}\mbox{ M}_{\bigodot}\mbox{ yr}^{-1}$,
which would be $2.2\times10^{-6}\mbox{ M}_{\bigodot}\mbox{ yr}^{-1}$
after correcting for our model stellar mass of $0.1\mbox{ M}_{\bigodot}$,
similar to the SED model prediction of $1.3\times10^{-6}\mbox{ M}_{\bigodot}\mbox{ yr}^{-1}$.
However, the cavity opening angle and disk mass for our models are
consistently smaller than those found by Stark et al. (2006).

If we again use the ratio $\mbox{A}_{V}/\tau_{SiO}=16.6\pm2.1$ from
Rieke \& Lebofsky (1985), we can use our measured optical depth $\tau_{SiO}=1.52\pm0.06$
to estimate the extinction toward DG Tau B to be $\mbox{A}_{V}=25.2\pm3.3\mbox{ mag}$.
Watson et al. (2004) also used this absorption feature to estimate
an extinction of 30 mag, which they associate with the circumstellar
disk. The best fit SED model shows a somewhat higher circumstellar
extinction of $\mbox{A}_{V}=36.1\mbox{ mag}$, but the circumstellar
extinction ranges from 26.6 to $\mbox{42 mag}$ for the disks within
the $\chi^{2}$ cut-off, comparable to our extinction estimate of
$25.2\pm3.3\mbox{ mag}$.

While the model results for cavity opening angle and disk mass are
small, the other parameters appear to be comparable to previous results
and estimates. Although the SED fitter returns circumstellar extinctions
that can be higher than previous estimates, the estimates are within
range of those disks in the $\chi^{2}$ cut-off.

\subsubsection{VV CrA}

While the IRC dominated the system flux at wavelengths >2.2 \textgreek{m}m
in the past, it has been progressively fading and is now fainter than
the primary at all wavelengths (see Table 8). We can thus only use
data taken within a relatively short period of time for modeling the
IRC. The Primary, on the other hand, has remained constant. Regular
observations by the the All Sky Automated Survey (Pojmanski 2002)
show the visually dominated primary has a mean Vmag=$13.8\pm0.3$ from
the beginning of 2001 through the end of 2009 without any trend. 

For the Primary fluxes at optical wavelengths, we use the UBVri magnitude
ratios reported in Marraco \& Rydgren (1981). Reipurth and Zinnecker
(1993) could not extract a Gunn z magnitude ratio for VV CrA because
the Primary was saturated. As the largest ratio they quote is 2\%,
and because the IRC is even weaker toward optical wavelengths relative
to the Primary (Koresko et al. 1997), we constrain the IRC flux to
be less than 2\% of the Primary for all wavelengths less than 0.91
\textgreek{m}m.

For the near-infrared magnitudes of the primary, we use flux values
reported by Prato et al. (2003) with observations taken in 1996 and
1997. Again assuming the primary does not vary, we then use relative
ratios found with CRIRES spectra reported in S09 to get a K-band IRC
flux. To get far-infrared information, we use ISO observations which
were taken when the IRC dominated the system flux at those wavelengths.
Because the IRC has faded while the Primary has remained constant,
we use the far-infrared magnitudes as upper limits for both the Primary
and IRC. 

For all models of VV CrA, we constrained the distance to be 120-160
pc away (Casey et al. 1998; de Zeeuw et al. 1999). VV CrA has been
classified as a late K-type (Appenzeller et al. 1986; Prato et al.
2003), so we also constrained ourselves to those models with stellar
temperatures $\lesssim$5,000 K. In order to model the two disk system
scenarios proposed by S09, we modeled the IRC twice with two interstellar
extinction ranges for the two cases. 

For the Case A structure of VV CrA, where the gas absorption is in
the Primary disk, the Primary disk will appear as extinction in the
line of sight to the IRC disk rather than a part of the IRC disk.
The SED model fitter uses the KMH model of dust grain sizes described
in Kim et al. (1994) to model the disk and interstellar extinction,
with a modification at wavelengths 1.25-8 \textgreek{m}m for the interstellar
extinction model (Indebetouw et al. 2005). Because the SED of the
IRC is not highly constrained in this wavelength region, we approximate
the Primary disk as having a grain size distribution similar to the
ISM, and thus model it as part of the interstellar extinction up to
an estimated 45 mag. For this SED fitting, the number of models returned
is very large for our $\chi^{2}$ cut-off, so we lower the cut-off
to $\chi_{best}^{2}-\chi^{2}\le3\mbox{N}$ so we are not overestimating
the parameters (Robitaille et al. 2007). The result for the best fit
seen in Fig. 11 is given in Table 9. The extinction that we would
associate with the Primary disk (the interstellar extinction) is 28.6
mag while the other models within the $\chi^{2}$ cut-off have a range
24.3 to 30.2 mag. Our estimate of $22.4\pm3\mbox{ mag}$ is at the
lower end of the model extinction range, similar to what was found
for DG Tau B.

For modeling S09 Case B, if the CO gas is only in the IRC disk, we
want to use the interstellar extinction seen toward the Primary as
an estimate for the IRC, so we first model the Primary (see Fig. 11).
The returned models have interstellar extinction $\mbox{A}_{V}$<1
mag, with the best fit model showing 0 mag. We compare this extinction
to the column density measured with the N($^{12}$CO)=$2.6\pm0.4\times10^{15}\mbox{ cm}^{-2}$
measured toward the Primary. The average N(CO)/$\mbox{A}_{V}=1.4\times10^{17}\mbox{ cm}^{-2}\mbox{ mag}^{-1}$
for the ISM (Rettig et al. 2006) would suggest a small $\mbox{A}_{V}\sim0.02\mbox{ mag}$,
showing this interstellar extinction range is accurate. We therefore
constrain the models for the IRC to have an interstellar extinction
A$_{V}$<1 mag. The resulting best fit seen in Fig. 11 has circumstellar
extinction of 43.4 mag, and the lowest extinction in the $\chi^{2}$
cut-off is 35.9 mag, much greater than the estimate of $22.4\pm3\mbox{ mag}$.

Thus, modeling the SED of the IRC for Case A, where the Primary disk
is the main source of extinction for the IRC, results in extinction
estimates that are more constistent with extinction estimates using
the optical depth $\tau_{SiO}$, suggesting this is a more likely
scenario.

\subsection{Gas and Dust}

The gas and dust in disks originate from the surrounding ISM, so if
we compare the dust extinction and CO abundance in the disk to the
ISM, we can investigate dust settling (Brittain et al. 2005). A change
in dust-to-gas ratio in the line of sight to a central emitting source
is largely due to the fact that the two components are not well mixed
throughout the disk. Models in Rettig et al. (2006) show that turbulence
in the disk will push dust away from the midplane, and the settling
velocity increases with grain size due to friction with the gas. This
results in a stratification with larger grains closer to the midplane,
and more grain growth leads to a higher gas/dust ratio above the midplane.
Further, the large gas/dust ratios require grain growth as well as
dust settling. We adopt the ratio used by Rettig et al. (2006) $\Delta=[\mbox{N(CO)/A}_{V}]_{disk}/[\mbox{N(CO)/A}_{V}]_{interstellar}$,
where $\mbox{[N(CO)/A}_{V}]_{interstellar}=1.4\times10^{17}\mbox{ cm}^{-2}\mbox{ mag}^{-1}$.
Rettig et al. (2006) probed the gas/dust ratio in the line of sight
to the central emitting source of four disks and found the $\Delta$-ratio
ranges from 1.8 for edge-on disks to $>\mbox{8}$ for face-on disks,
and we use these values as a starting point of comparison for DG Tau
B and VV CrA.

Rettig et al. (2006) uses A$_{V}$ estimates from JHK-band photometry,
so the gas and dust are being probed at similar wavelengths. Our extinction
estimates are taken from these wavelengths as well as the rest of
the SED. However, while the near-infrared spectrum of DG Tau B and
the VV CrA Primary are well-constrained, this is not the case for
the VV CrA IRC. Thus, this should be taken as a first estimate for
the IRC.

\subsubsection{DG Tau B}

We use our $^{12}$CO column density of $2.42\pm0.04\times10^{19}\mbox{ cm}^{-2}$
and circumstellar extinction, estimated using the $\tau_{SiO}$, of
$25.2\pm3.3\mbox{ mag}$ to find $\Delta=6.86\pm0.9$. This is significantly
larger than the ratios found toward other edge-on disks in Rettig
et al. (2006), which were $\sim$1.8. If we use the extinction found
with the SED modeling, $\mbox{36.1 mag}$, the $\Delta$-ratio would
still be 4.8. Further, we are using the column density estimated from
the $^{13}$CO column density and the mean local ISM isotopic ratio
(Wilson 1999). If the $^{12}$CO is self-shielding, as was found in
the disks in S09, the N($^{12}$CO)/N($^{13}$CO) would increase from
the ISM, and the $\Delta$-ratio would only be larger. This indicates
a larger grain growth and settling in DG Tau B than seen toward the
disks in Rettig et al. (2006). 

For comparison, Lahuis et al. (2006) found $\tau_{SiO}=0.86$ (A$_{V}=16.6\tau_{SiO}=14.2\mbox{ mag}$) and N(CO)=$2\times10^{18}\mbox{ cm}^{-2}$, giving $\Delta=1.0$ for IRS 46.  For GV Tau N, Bowey \& Adamson (2001) found $\tau_{SiO}=1.7$ and Gibb et al. (2007) found N(CO)=$5.9\times10^{18}\mbox{ cm}^{-1}$, giving $\Delta=1.5$.  IRS 46 and GV Tau N thus have gas/dust ratios more similar to the ISM, as expected for edge-on disks, further emphasizing the high $\Delta$-ratio found in DG Tau B.  This also suggests the line of sight in DG Tau B is not probing deeper in the disk atmosphere than IRS 46 and GV Tau N, nearer where the dust is settled, as is suggested by the $\tau_{SiO}/\tau_{CO_{2}}$ ratio.  Rather, the higher $\tau_{SiO}/\tau_{CO_{2}}$ ratios in IRS 46 and GV Tau N may just be due to having lines of sight that view more silicate in the warmer regions of the disk, where the CO$_2$ ice has sublimated, than for DG Tau B.

\subsubsection{VV CrA}

The $\Delta$-ratios in Rettig et al. (2006) are for probing the gas/dust
ratio in the radial structure toward the central emitting source.
This method can therefore not be used to test the Case A structure
for VV CrA, as the geometry in this scenario would be probing the
vertical structure of the Primary disk. For Case B, the VV CrA IRC
was found by S09 to have a hot $^{12}$CO gas component with column
density $7.15\pm0.17\times10^{18}\mbox{ cm}^{-2}$ and cold $^{12}$CO
gas of $>1.40\pm0.07\times10^{18}\mbox{ cm}^{-2}$. By adding these
column densities, and using the A$_{V}$ estimate of $22.4\pm3$ found
with the silicate absorption, we find $\Delta\ge$2.74. Comparing
to the other sources in Rettig et al. (2006), this is comparable to
a disk with inclination $\sim60^{\circ}$.

\subsection{Velocity Shift in DG Tau B}

The warm gases toward IRS 46 and GV Tau N show velocity shifts that suggest they are not in simple Keplerian rotation in the disk.  IRS 46 shows $^{12}$CO and HCN blueshifted $\sim24\mbox{ km s}^{-1}$ from the quiescent cloud (Lahuis et al. 2006), and GV Tau N shows warm HCN redshifted $\sim13\mbox{ km s}^{-1}$ from the stellar velocity (Doppmann et al. 2008).  Similarly, DG Tau B shows a shift that raises the question of where the gas resides.

The molecular cloud surrounding DG Tau B has a velocity $\mbox{v}_{LSR}=6.5\mbox{ km s}^{-1}$ (Mitchell et al. 1994), so we would expect a similar velocity for DG Tau B (Herbig 1977).  However, the $^{13}$CO and high-J $^{12}$CO gas seen toward DG Tau B have an excess redshift of $\sim6-7\mbox{ km s}^{-1}$, so the gas does not appear to be in simple Keplerian rotation.  The gas is not likely associated with a disk wind, which would result in a blue shift, or with the optical jets as they have velocities $>50\mbox{ km s}^{-1}$ and $<-90\mbox{ km s}^{-1}$ relative to the cloud velocity (Eisl\"{o}ffel \& Mundt 1998).  A molecular outflow was found to have a blue (v$_{LSR}$=-3.8 to 5.1 $\mbox{km s}^{-1}$) and red (7.5 to 10.9$\mbox{ km s}^{-1}$; Mitchell et al. 1997) component, but our observed shift is outside these bounds.  While the CO velocity shift is near the extreme end of the red outflow, JHK-band images show the main source of continuum is southeast of the dark lane (Padgett et al. 1999), in the direction of the blue outflow component.  The molecular outflow does encase the surrounding cloud velocity, indicating the DG Tau B systemic velocity is likely similar to the surrounding cloud.

One explanation for the observed shift could be infalling gas.  Assuming the stellar mass of $0.1
\mbox{ M}_{\bigodot}$ from our SED model fitting, the velocity of the infalling gas $\mbox{v}_r=
(2GM_\star/r)^{1/2}$ would be $6\mbox{ km s}^{-1}$ at $r=5\mbox{ AU}$.  An envelope 
infall rate of $1.3\times10^{-6}\mbox{ M}_{\bigodot}\mbox{ yr}^{-1}$ would imply a density $n_
{H}=\dot{M}/(2 \pi r^{2} \mbox{v} m_{H})=2.3\times10^9\mbox{ cm}^{-3}$.  Assuming a CO/H ratio 
of $\sim3\times10^{-4}$ and a column length of 5 AU, we would estimate a column density of $5
\times10^{19}\mbox{ cm}^{-2}$, similar our measured $2.4\times10^{19}\mbox{ cm}^{-2}$.
At this radius, the accretion radiation from an infalling envelope has been modeled to heat 
the gas to $\sim300\mbox{ K}$ for L$\ge20\mbox{ L}_{\bigodot}$ (see Fig. 4 in Ceccarelli et al. 
1996).  However, the total luminosity of DG Tau B is quite small compared to those models, found 
previously to range between 0.2 and 2.5 L$_{\bigodot}$ (Stark et al. 2006; Furlan et al. 2008), and 
between 0.7 and 1.5 L$_{\bigodot}$ by our SED model fits.  If we were to increase the stellar mass to $0.5\mbox{ M}_{\bigodot}$, this would push a $6\mbox{ km s}^{-1}$ infall out to 25 AU where even an accretion radiation of 65 L$_{\bigodot}$ would result in gas temperatures $<200\mbox{ K}$.  Also, if the CO gas is associated with 
an infalling envelope, we would expect the gas/dust ratio to be near the ISM value, contrary to our 
findings.  The absorption is thus unlikely to be associated with an infalling envelope.

Another possible explanation for the observed velocity shift could be the CO is in the disk of a non-axisymmetric system.  The main M-band continuum source is likely the disk which could be non-axisymmetric, possibly due to having a close stellar or planetary companion, hot spots in the disk, or some similar phenomenon.  Also, the absorbing gas may be non-axisymmetric, as would be the case for an asymmetric disk atmosphere.  In either case, the absorption would vary in velocity but average out over time to the systemic velocity.  Although observations that show such velocity variation would be needed to support this hypothesis, it currently appears to be the most likely scenario.

\subsection{Conclusion}

While the Spitzer Infrared Spectrograph was an excellent instrument
to detect molecular signatures in circumstellar disks, the detection
of organic molecules in absorption was rare. DG Tau B is one disk
that showed molecular gas in absorption. While we were able to directly
detect CO$_{2}$, we only found upper limits for C$_{2}$H$_{2}$
and HCN, the other species found toward IRS 46 with the Spitzer IRS
(Lahuis et al. 2006). This could simply be a function of lower gas
abundances in the disk, or this could be due to probing a slightly
different region of the disk where C$_{2}$H$_{2}$ and HCN have lower
abundances, as is suggested by the gas column densities and gas/ice
ratios. Although we are viewing an edge-on disk like IRS 46 and GV
Tau N, this could imply molecular absorption is highly sensitive to
even moderate changes in the line of sight in the disk, which may
account for the rarity of such detections even in edge-on disks. Further,
the large CO column density to extinction ratio indicates the disk
is likely showing a large amount of grain growth and dust settling
when compared to other disks (Rettig et al. 2006).  Finally, while the velocity shift of the 
absorbing CO gas is difficult to explain, the most likely scenario currently
appears to be the CO is in the disk while the main continuum emitting source
or the absorbing gas is non-axisymmetric.

While the Spitzer IRS spectrum of VV CrA shows only silicate and CO$_{2}$
ice absorption, high-resolution spectroscopy reveals the disk has
a complicated and changing near-IR spectrum. Smith et al. (2009) proposed
the absorbing CO gas seen toward the IRC is located (A) within the
Primary disk in the line of sight to the IRC, or (B) within the disk
around the IRC. To test these cases, we used T-ReCS images to separate
the fluxes of the VV CrA components in the IRS spectrum, allowing
us to determine a dust extinction estimate for the IRC. We then modeled
the SED for the IRC with two interstellar extinction ranges, simulating
the two cases proposed by S09, and found the disk models for Case
A return circumstellar extinction magnitudes more consistent with
our extinction estimate. This supports the view by Smith et al. (2009)
that this is a more likely case.

\acknowledgements{We would like to gratefully thank Thomas Robitaille, Thorsten Ratzka, John Lacy, and the anonymous reviewer for their constructive help, comments and suggestions. Support for
this work was provided by the National Science Foundation under Grant
No. AST-0708074, and by NASA through contract RSA No. 1346810, issued
by JPL/Caltech. This work is based on observations made with the Spitzer
Space Telescope, Gemini Observatory, and W.M. Keck Observatory. The
Spitzer Space Telescope is operated by the Jet Propulsion Laboratory,
California Institute of Technology under a contract with NASA. The
Gemini Observatory is operated by the Association of Universities
for Research in Astronomy, Inc., under a cooperative agreement with
the NSF on behalf of the Gemini partnership: the National Science
Foundation (United States), the Science and Technology Facilities
Council (United Kingdom), the National Research Council (Canada),
CONICYT (Chile), the Australian Research Council (Australia), Minist\'{e}rio
da Ci\^{e}ncia e Tecnologia (Brazil) and Ministerio de Ciencia, Tecnolog\'{i}a
e Innovaci\'{o}n Productiva (Argentina). The W.M. Keck Observatory is
operated as a scientific partnership among the California Institute
of Technology, the University of California and the National Aeronautics
and Space Administration. The Observatory was made possible by the
generous financial support of the W.M. Keck Foundation. Basic research in infrared 
astronomy at the Naval Research Laboratory is supported by 6.1 base funding.}




\clearpage

\begin{table}
\caption{Detailed list of observations taken for this work. }

\begin{tabular}{ccccccc}
\hline 
Target & Date & Instrument & Mode & Setting & Total Int. Time (s) & Standard Star\tabularnewline
\hline
\hline 
VV CrA & 2008 Nov. 18 & Spitzer IRS & Spectrum & SL1, SL2 & 151 & \tabularnewline
 &  &  &  & SH & 151 & HR 6688\tabularnewline
 & 2009 April 6 & Phoenix & Spectrum & 4.66 \textgreek{m}m & 1920 & HD 209952\tabularnewline
 & 2009 June 8 & T-ReCS & Image & %
Si-1, 7.37 \textgreek{m}m & %
319 & HD 181109\tabularnewline
 &  &  &  & %
Si-2, 8.74 \textgreek{m}m & %
203 & HD 181109\tabularnewline
 &  &  &  & %
Si-4, 10.38 \textgreek{m}m & %
203 & HD 181109\tabularnewline
 &  &  &  & %
Si-6, 12.33 \textgreek{m}m & %
203 & HD 181109\tabularnewline
 &  &  &  & %
Qa, 18.3 \textgreek{m}m & %
261 & HD 196171 \tabularnewline
 &  &  &  & %
Qb, 24.56 \textgreek{m}m & %
319 & HD 196171 \tabularnewline
\hline
DG Tau B & 2008 Nov. 5 & Spitzer IRS & Spectrum & SL1, SL2 & 151 & \tabularnewline
 &  &  &  & SH & 151 & HR 6688\tabularnewline
 & 2008 Nov. 16 & NIRSPEC & Spectrum & 4.6 and 4.9 \textgreek{m}m & 3360 & HR 1641\tabularnewline
\hline
\end{tabular}
\end{table}

\clearpage{}

\begin{table}
\caption{Measured equivalent widths and doppler shifts of the CO fundamental
absorption lines used in measuring column density and temperatures
for DG Tau B.}

\begin{tabular}{ccc}
\hline 
\multicolumn{3}{c}{DG Tau B}\tabularnewline
\hline 
Line ID & Equivalent Width (cm$^{-1}$) & Doppler Shift (v${}_{LSR}$, km s$^{-1}$)\tabularnewline
\hline
\hline 
$^{12}$CO P(25) & 0.0458$\pm$0.003 & 14.7$\pm$1.3\tabularnewline
$^{12}$CO P(26) & 0.0451$\pm$0.002 & 13.2$\pm$0.6\tabularnewline
$^{12}$CO P(27) & 0.0423$\pm$0.002 & 13.1$\pm$0.6\tabularnewline
$^{12}$CO P(30) & 0.0282$\pm$0.001 & 13.9$\pm$0.8\tabularnewline
$^{12}$CO R(0) & 0.111$\pm$0.003 & 9.5$\pm$0.3\tabularnewline
$^{12}$CO R(1) & 0.120$\pm$0.004 & 9.7$\pm$0.3\tabularnewline
$^{12}$CO R(2) & 0.117$\pm$0.004 & 9.6$\pm$0.3\tabularnewline
$^{12}$CO R(3) & 0.118$\pm$0.004 & 8.8$\pm$0.3\tabularnewline
$^{12}$CO R(5) & 0.125$\pm$0.01 & 8.9$\pm$0.5\tabularnewline
$^{12}$CO R(6) & 0.127$\pm$0.007 & 8.6$\pm$0.4\tabularnewline
$^{12}$CO R(7) & 0.134$\pm$0.007 & 9.6$\pm$0.4\tabularnewline
$^{12}$CO R(8) & 0.114$\pm$0.005 & 8.1$\pm$0.4\tabularnewline
$^{12}$CO R(9) & 0.136$\pm$0.009 & 8.3$\pm$0.5\tabularnewline
$^{13}$CO R(15) & 0.0175$\pm$0.0006 & 12.8$\pm$1.0\tabularnewline
$^{13}$CO R(16) & 0.0180$\pm$0.0006 & 12.3$\pm$0.7\tabularnewline
$^{13}$CO R(17) & 0.0133$\pm$0.005 & 10.7$\pm$1.1\tabularnewline
$^{13}$CO R(18) & 0.0079$\pm$0.0003 & 14.0$\pm$2.2\tabularnewline
$^{13}$CO R(19) & 0.0091$\pm$0.0004 & 10.6$\pm$2.2\tabularnewline
\hline
\end{tabular}%

\end{table}

\clearpage{}

\begin{table}
\caption{Measured parameters for molecular absorption toward DG Tau B and
VV CrA.}

\begin{tabular}{cccccc}
\hline 
Target & Molecule & T (K) & N ($10^{17}$ cm$^{-2}$) & v${}_{LSR}$ (km s$^{-1}$) & FWHM (km s$^{-1}$) $^{a}$\tabularnewline
\hline
\hline 
DG Tau B & high-J $^{12}$CO$^{b}$ & 370$\pm$20 & $\mbox{240}\pm20$ & 13.4$\pm$0.4 & 13.7$\pm$0.4\tabularnewline
 & low-J $^{12}$CO & - & - & 9.3$\pm$0.6 & 20.0$\pm$0.4\tabularnewline
 & $^{13}$CO & 288$\pm$7 & $\mbox{3.50}\pm\mbox{.06}$ & 12.0$\pm$0.5 & 14.3$\pm$0.4\tabularnewline
 & CO$_{2}$ & 300 & 0.29 & - & 2.1$^{c}$\tabularnewline
VV CrA Primary & $^{12}$CO & 29$\pm$4 & $\mbox{0.026}\pm0.004$ & 4.88$\pm$0.4 & 5.35$\pm$0.2\tabularnewline
VV CrA IRC & $^{12}$CO & - & - & 0.78$\pm$0.05 & 14.8$\pm$0.2\tabularnewline
 & $^{13}$CO & - & - & 1.75$\pm$0.3 & 10.2$\pm$0.8\tabularnewline
\hline
\end{tabular}

$^{a}$Line widths are of convolved spectra.

$^{b}$Assuming $^{12}$CO/$^{13}$CO abundance ratio of 69$\pm$6
(Wilson 1999).

$^{c}$Doppler b-value, see text.
\end{table}
\clearpage{}

\begin{table}
\caption{Measured equivalent widths and doppler shifts of the CO fundamental
absorption lines for VV CrA.}

\begin{tabular}{cccc}
\hline 
\multicolumn{2}{c}{} & \tabularnewline
\hline 
Target & Line ID & Equivalent Width (cm$^{-1}$) & Doppler Shift (v${}_{LSR}$, km s$^{-1}$)\tabularnewline
\hline
\hline 
IRC & $^{12}$CO R(3) & 0.102$\pm$0.01 & 1.1$\pm$0.1\tabularnewline
 & $^{12}$CO R(4) & 0.105$\pm$0.02 & 0.3$\pm$0.1\tabularnewline
 & $^{12}$CO R(5) & 0.099$\pm$0.005 & 0.8$\pm$0.06\tabularnewline
 & $^{13}$CO R(18) & 0.0106$\pm$0.0008 & 1.6$\pm$0.4\tabularnewline
 & $^{13}$CO R(19) & 0.0055$\pm$0.0004 & 1.8$\pm$0.3\tabularnewline
Primary & $^{12}$CO R(3) & 0.0127$\pm$0.0005 & 5.2$\pm$0.1\tabularnewline
 & $^{12}$CO R(4) & 0.0115$\pm$0.0004 & 4.5$\pm$0.1\tabularnewline
 & $^{12}$CO R(5) & 0.0013$\pm$0.00005 & 5.4$\pm$0.4\tabularnewline
\hline
\end{tabular}
\end{table}
\clearpage{}

\begin{table}

\caption{Fluxes for the VV CrA Primary and IRC as measured with T-ReCS imaging.}

\begin{tabular}{ccc}
\hline 
 & Primary & IRC\tabularnewline
\cline{2-3} 
Filter & Flux (Jy) & Flux (Jy)\tabularnewline
\hline
\hline 
Si-1, 7.73 \textgreek{m}m & 13.7$\pm$0.7 & 7.0$\pm$0.4\tabularnewline
Si-2, 8.74 \textgreek{m}m & 14.4$\pm$0.2 & 4.5$\pm$0.1\tabularnewline
Si-4, 10.38 \textgreek{m}m & 15.0$\pm$0.3 & 3.3$\pm$0.1\tabularnewline
Si-6, 12.33 \textgreek{m}m & 20.5$\pm$0.7 & 8.7$\pm$0.3\tabularnewline
Qa, 18.3 \textgreek{m}m & 28.4$\pm$1.6 & 11.1$\pm$0.8\tabularnewline
Qb, 24.56 \textgreek{m}m & 35.0$\pm$17.8 & 25.1$\pm$13.5\tabularnewline
\hline
\end{tabular}

\end{table}
\clearpage{}

\begin{table}
\caption{Measured column densities and ratios for DG Tau B compared to IRS
46 (Lahuis et al. 2006)}

\begin{tabular}{ccc}
\hline 
Abundance Ratios & DG Tau B & IRS 46\tabularnewline
\hline
\hline 
N(CO) ($10^{18}\mbox{ cm}^{-2}$) & 24.2$^{a}$ & 2\tabularnewline
N(HCN)/N(CO) & < 0.00026 & 0.025\tabularnewline
N(C$_{2}$H$_{2}$)/N(CO) & < 0.00023 & 0.015\tabularnewline
N(CO$_{2}$ gas)/N(CO) & 0.0012 & 0.05\tabularnewline
N(CO$_{2}$ ice)/N(CO)$^{b}$ & 0.0022 & 0.12\tabularnewline
$\tau_{SiO}/\tau_{CO_{2}ice}$ & 6.6 & 8.6\tabularnewline
\hline
\end{tabular}

$^{a}$Using $^{13}$CO column density and isotopic ratio
in local ISM, see text.

$^{b}$CO$_{2}$ ice column densities are taken from Pontoppidan et
al. (2008)
\end{table}

\clearpage{}

\begin{table}
\caption{Estimated disk parameters from SED fitting of DG Tau B. Shown are the best fit and the (min, max) of the best fit models.}
\begin{tabular}{c|ccc}
\hline 
\multicolumn{1}{c}{Parameter} & DG Tau B & Literature & Reference\tabularnewline
\hline
\hline 
Circumstellar A$_{V}$ (mag) & 36.1 (26.6, 42)& $\sim30$ & 2\tabularnewline
Interstellar A$_{V}$ (mag) & 10.5 (6.5, 12.4) &  & \tabularnewline
Stellar Temperature (K) & 2540 (2540, 2790) & 3800$^{a}$ & 1\tabularnewline
Env. Infall Rate (10$^{-6}$ M$_{\bigodot}$ yr$^{-1}$) & 1.3 (0.8, 1.8)  & 2.5$^{b}$ & 1\tabularnewline
Disk Accretion Rate ($10^{-8}\mbox{ M}_{\bigodot}\mbox{ yr}^{-1}$) & 22 (0.4, 48) &  & \tabularnewline
Disk Mass (10$^{-4}$ M$_{\bigodot}$) & 4.4 (1.8, 185) & 400 & 1\tabularnewline
Disk Inner Radius (AU) & 0.056 (0.056, 0.44) & 0.035$^{a}$ & 1\tabularnewline
Cavity Opening Angle ( $^{\circ}$ ) & 3.3 (1.9, 6.2) & 30 & 1\tabularnewline
\hline
\end{tabular}

References.-- (1) Stark et al. (2006); (2) Watson et al. (2004)


$^{a}$Used as constants in Stark et al. (2006).

$^{b}$Corrected for model stellar mass, see text.
\end{table}
\clearpage{}

\begin{table}
\caption{VV CrA flux ratios (IRC / Primary) in literature.}
\begin{tabular}{ccccccc}
\hline 
J & H & K & L & M & Date Obs. & Reference\tabularnewline
\hline
\hline 
1.02 & 4.66 & 7.73 & 10.76 &  & 1987 June 6-8 & Chelli et al. 1995\tabularnewline
 &  & 0.515$\pm$0.004 &  &  & 1995 July 12 & K\"{o}hler et al. 2008\tabularnewline
0.103 & 0.299 & 0.946 &  &  & 1996 Jan. 7 & Prato et al. 2003\tabularnewline
0.017$\pm$0.0001 & 0.111$\pm$0.008 & 0.358$\pm$0.013 &  &  & 2001 July 6 & K\"{o}hler et al. 2008\tabularnewline
0.029 &  & 0.34$^{a}$ & 0.69$^{a}$ & 0.72$^{a}$ & 2003 July 9 & Ratzka et al. 2008\tabularnewline
 &  & 0.122$\pm$0.006$^{b}$ &  & 0.31$\pm$0.01$^{b}$ & 2007 Sept. 9 & This work: CRIRES\tabularnewline
 &  &  &  & 0.41$\pm$0.01$^{c}$ & 2009 April 6 & This work: Phoenix\tabularnewline
\hline
\end{tabular}%

$^{a}$The filters used are Ks, L', and M'.

$^{b}$Taken at 2.37 and 4.72 \textgreek{m}m.

$^{c}$Taken at 4.62 \textgreek{m}m.
\end{table}
\clearpage{}

%

\begin{table}
\caption{Estimated disk parameters from SED fitting of the VV CrA Primary and
for the IRC for Cases A and B. Shown are the best fit and the (min, max) of the best fit models.}
\begin{tabular}{c|ccc}
\hline 
\multicolumn{1}{c}{Parameter} & Primary & IRC: Case A & IRC: Case B\tabularnewline
\hline
\hline 
Circumstellar A$_{V}$ (mag) & 26.0 (20, 31) & 0.15 (0.0003, 0.15) & 43.4 (36, 43.4)\tabularnewline
Interstellar A$_{V}$ (mag) & 0 (0, 1.0) & 28.6 (24.4, 30) & 0 (0, 1.0)\tabularnewline
Stellar Temperature (K) & 3560 (3440, 3650) & 5180 (5070, 5410) & 3380 (3240, 3380)\tabularnewline
Env Infall Rate (10$^{-6}$ M$_{\bigodot}$ yr$^{-1}$) & 3.5 (2.5, 6.3) & 0.02 (0, 0.02) & 2.1(1.7, 2.1) \tabularnewline
Disk Accretion Rate ($10^{-8}\mbox{ M}_{\bigodot}\mbox{ yr}^{-1}$) & 6.7 (6.7, 200) & 4.1 (4.1, 6.1) & 4.1 (4.1, 150)\tabularnewline
Disk Mass (10$^{-3}$ M$_{\bigodot}$) & 1.5 (1.1, 7.3) & 26.7 (4.4, 80) & 7.5 (7.0, 7.5)\tabularnewline
Disk Inner Radius (AU) & 0.2 (0.15, 8.0) & 3.5 (2.2, 3.5) & 0.1 (0.1, 1.3)\tabularnewline
Cavity Opening Angle ( $^{\circ}$ ) & 5.9 (2.9, 7.7) & 55.5 (32, 55.5) & 1.4 (1.4, 2.0)\tabularnewline
\hline
\end{tabular}
\end{table}

\clearpage{}

\begin{figure}
\begin{center}
\includegraphics[angle=90,scale=0.48]{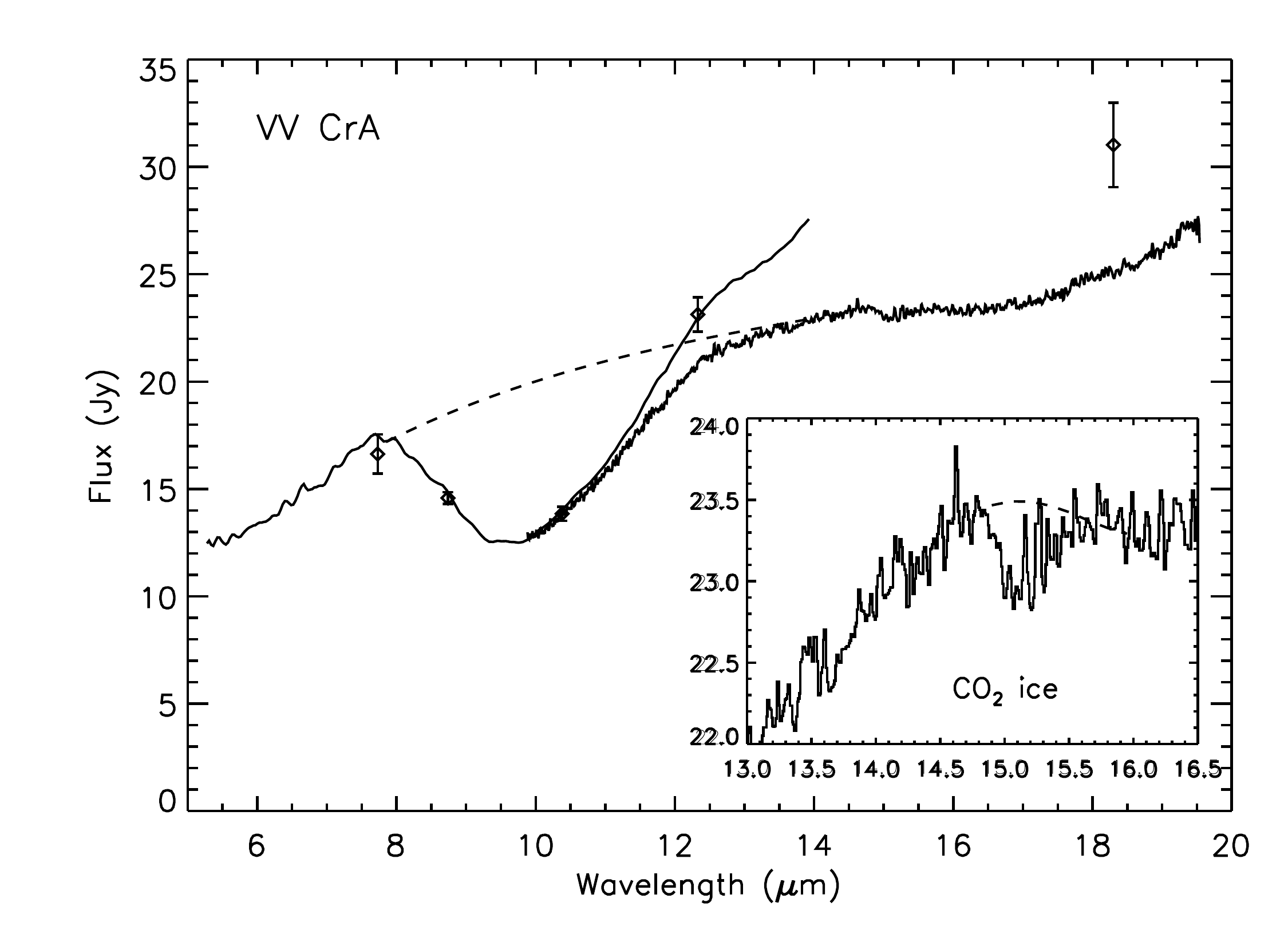}
\end{center}
\begin{center}
\includegraphics[angle=90,scale=0.48]{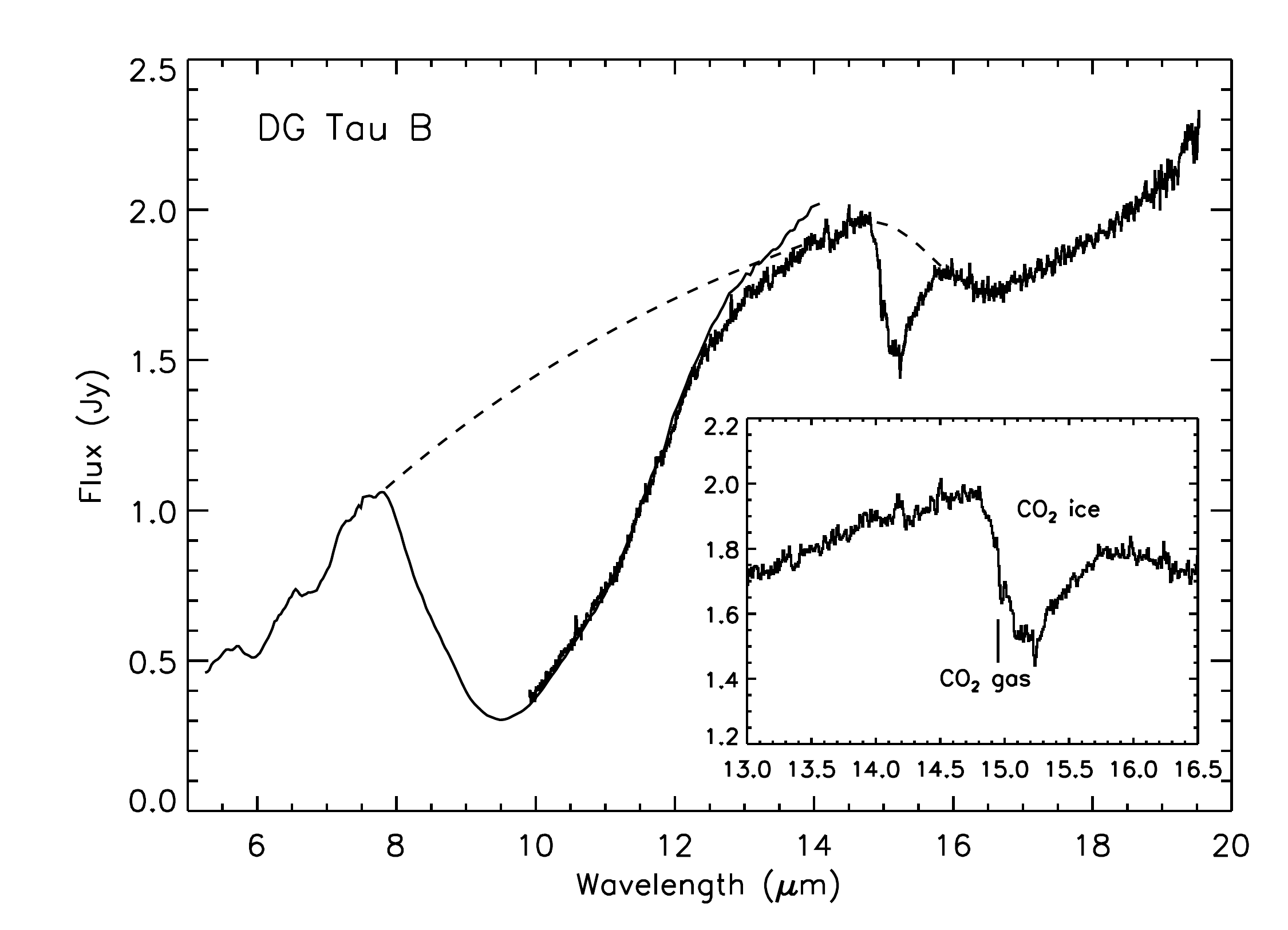}
\end{center}
\caption{The Spitzer IRS spectra of DG Tau B and VV CrA. DG Tau B displays
water (6 \textgreek{m}m) and {}``methanol'' (6.8 \textgreek{m}m)
ice absorption; both targets show amorphous silicates (8-12
and 15-19 \textgreek{m}m) and CO$_{2}$ ice (15.2 \textgreek{m}m)
absorption. Overplotted are the continua used to find the silicate and CO$_{2}$ ice peak optical depths (dashed lines; see text).  The VV CRA spectrum also shows the summed fluxes of the
binary components as found with T-ReCS (diamonds), with each point summing 70\%
of the Primary flux and 100\% of the IRC flux.}

\end{figure}

\clearpage{}%

\begin{figure}
\begin{center}
\includegraphics[angle=90,scale=0.6]{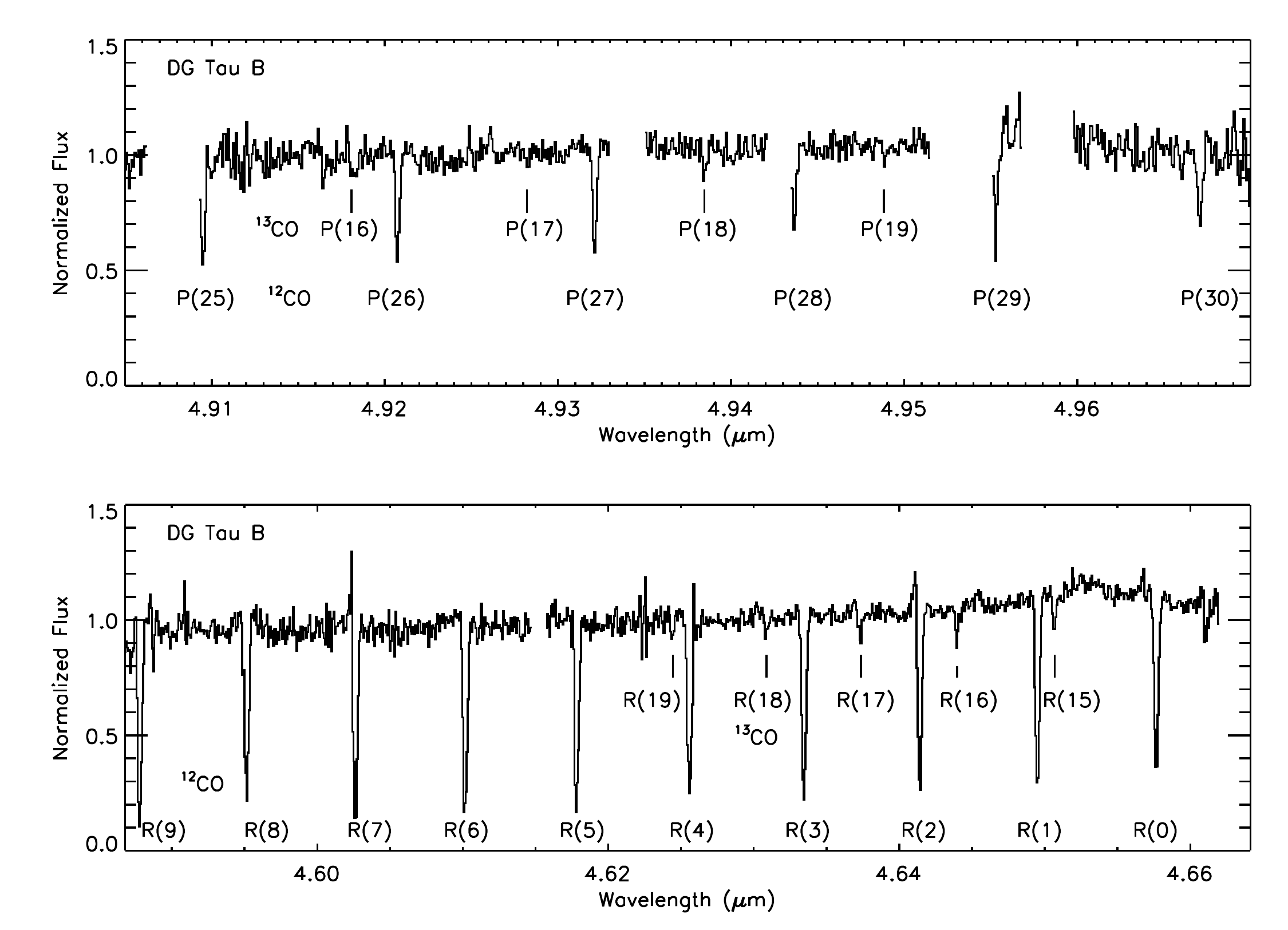}
\end{center}
\caption{The CO fundamental absorption spectra for DG Tau B in orders 15 (above)
and 16 (below) with detected transitions marked. }

\end{figure}

\clearpage{}%

\begin{sidewaysfigure}
\includegraphics[angle=90,scale=0.3]{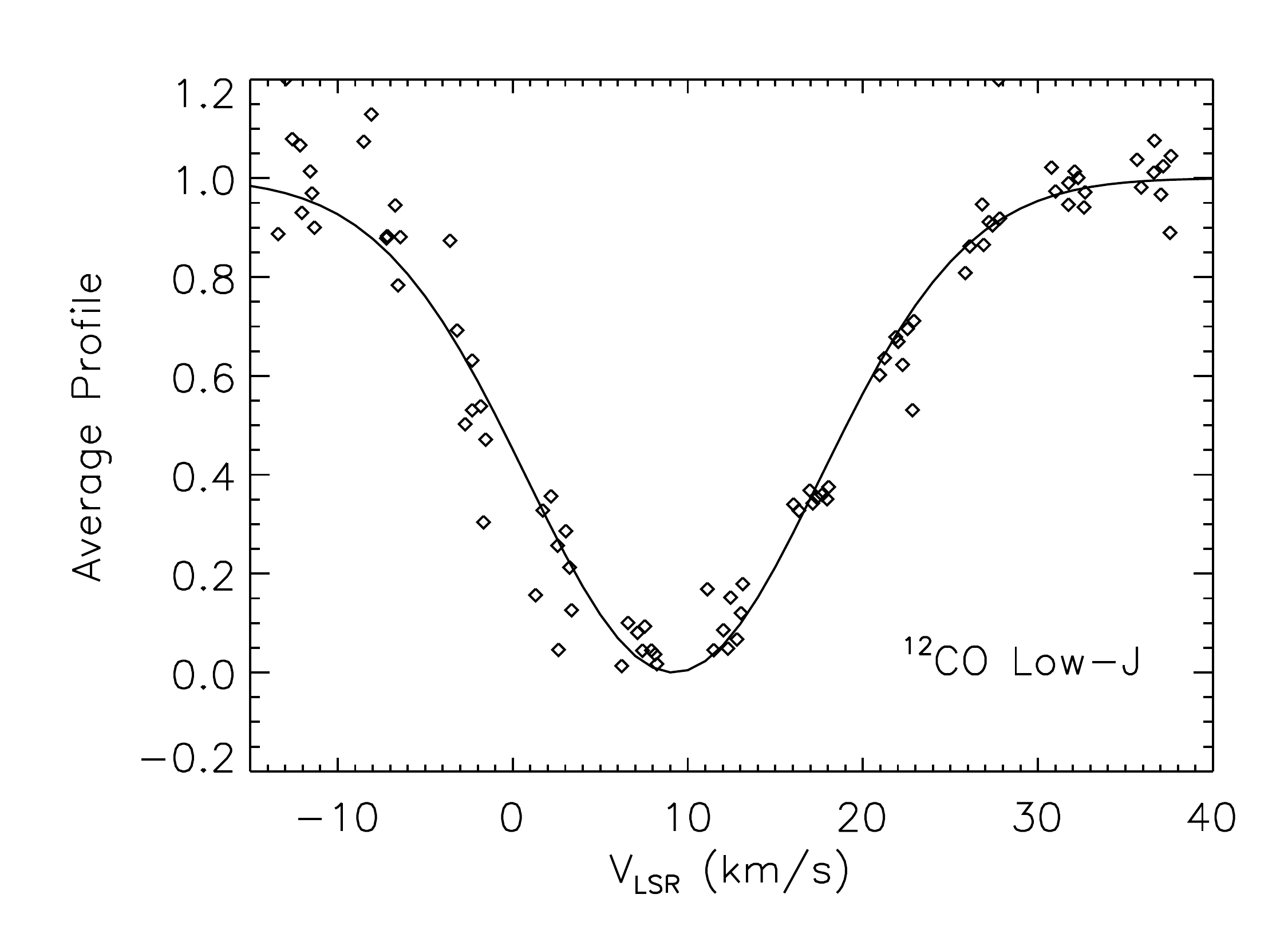}\includegraphics[angle=90,scale=0.3]{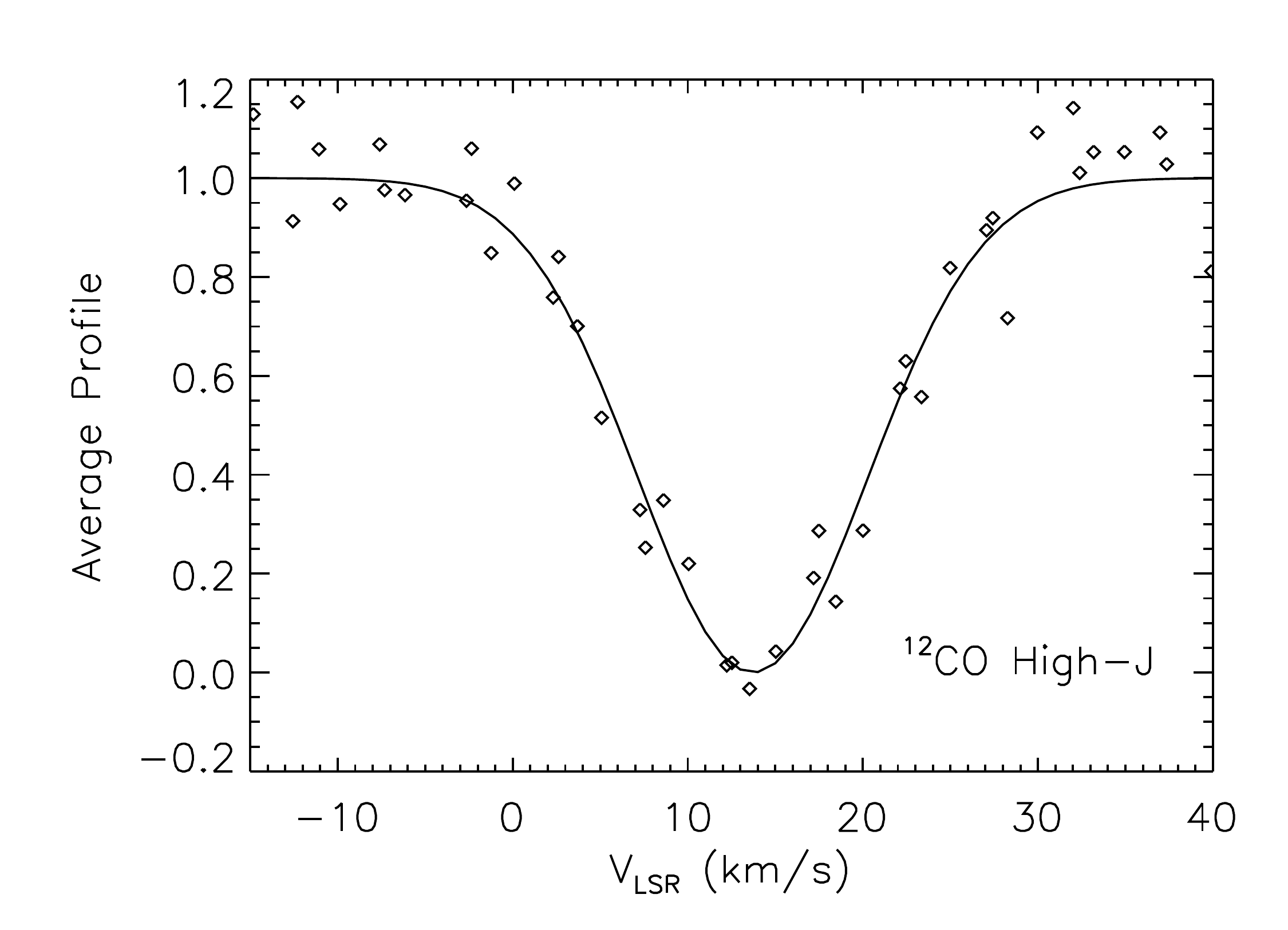}\includegraphics[angle=90,scale=0.3]{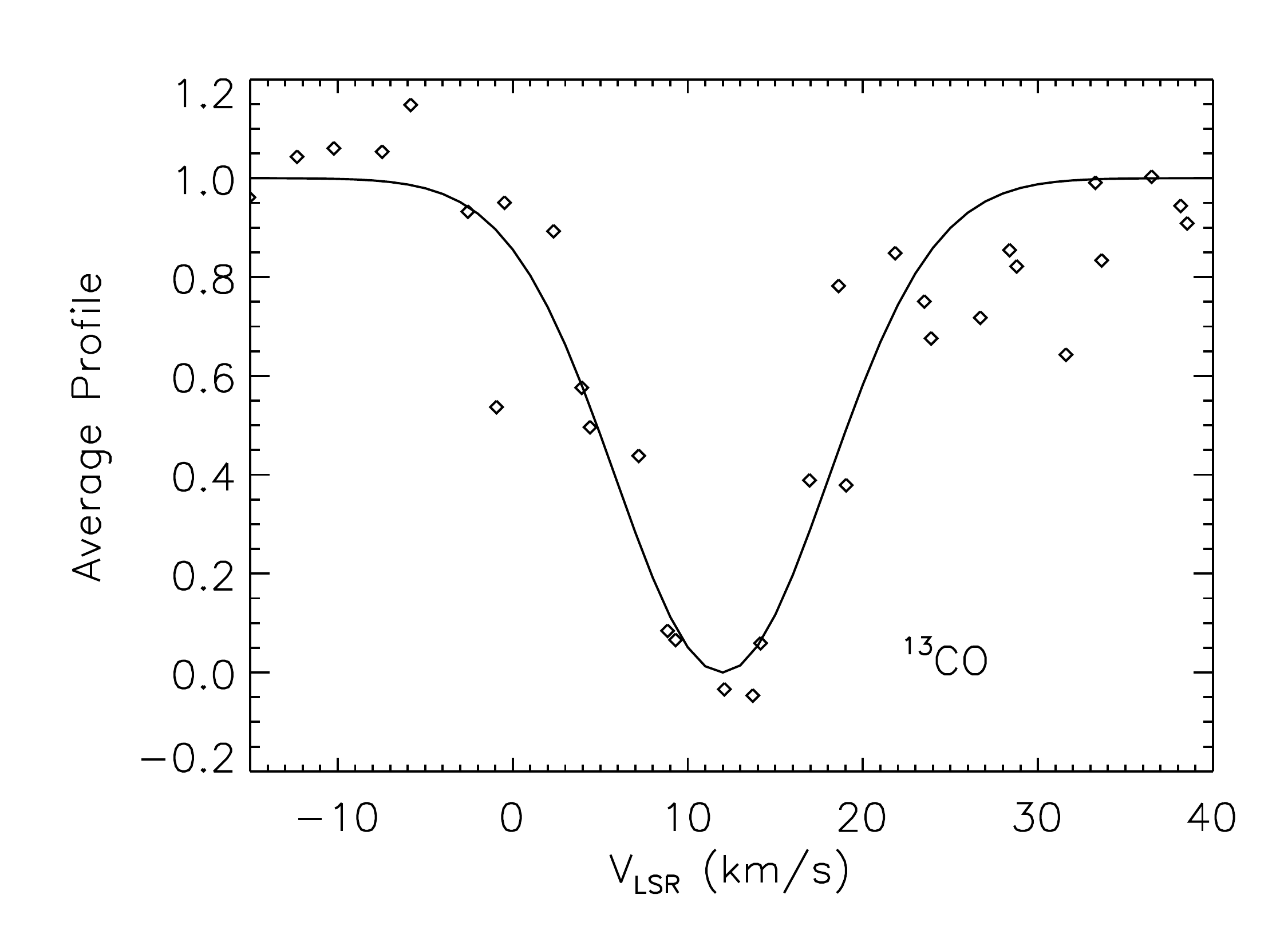}\caption{$^{12}$CO and $^{13}$CO line profiles for DG Tau B. It can be seen
the $^{12}$CO low-J R(0-7) transitions (left) have a different profile
from the $^{12}$CO high-J P(25-27, 30) (middle) and $^{13}$CO R(15-17)
transitions (right).}

\end{sidewaysfigure}

\clearpage{}%

\begin{figure}
\begin{center}
\includegraphics[trim=230 0 0 0,clip,angle=90,scale=0.6]{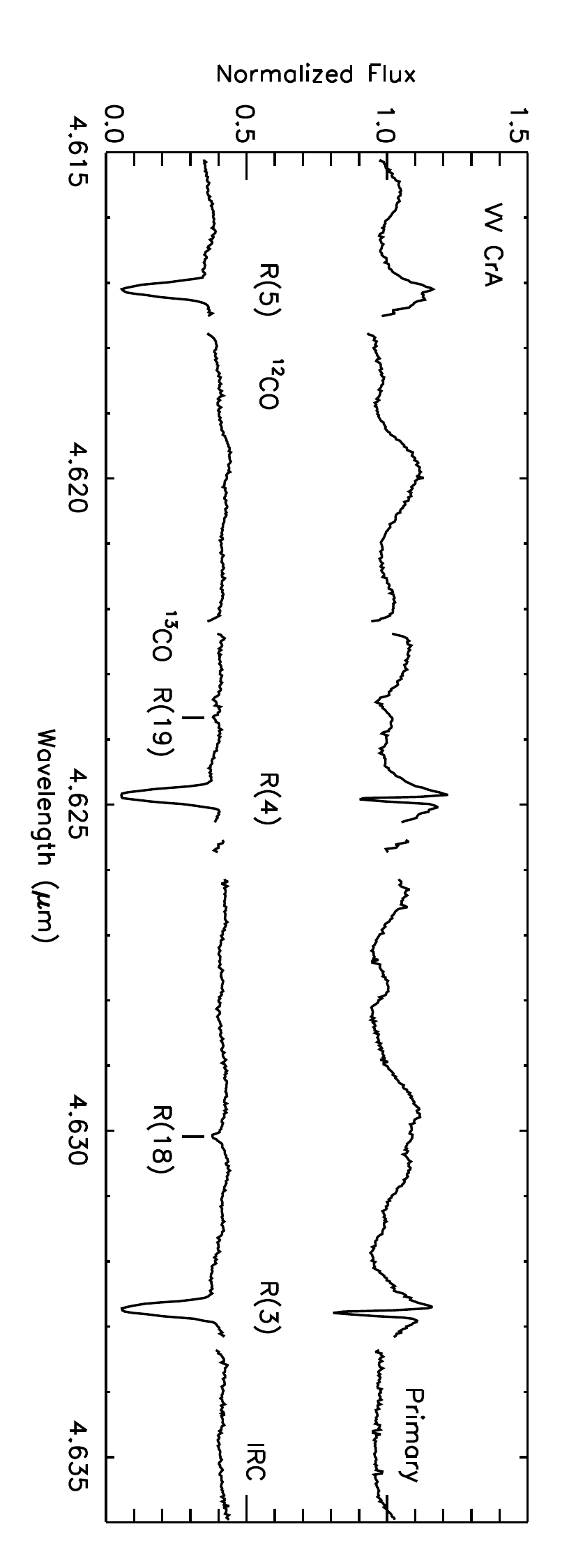}
\end{center}
\caption{The CO fundamental absorption spectra for the VV CrA binary components
with detected transitions marked. The primary (upper spectrum) and
IRC (lower spectrum) are normalized by the primary flux.}

\end{figure}

\clearpage{}%

\begin{figure}
\begin{center}
\includegraphics[angle=90,scale=0.4]{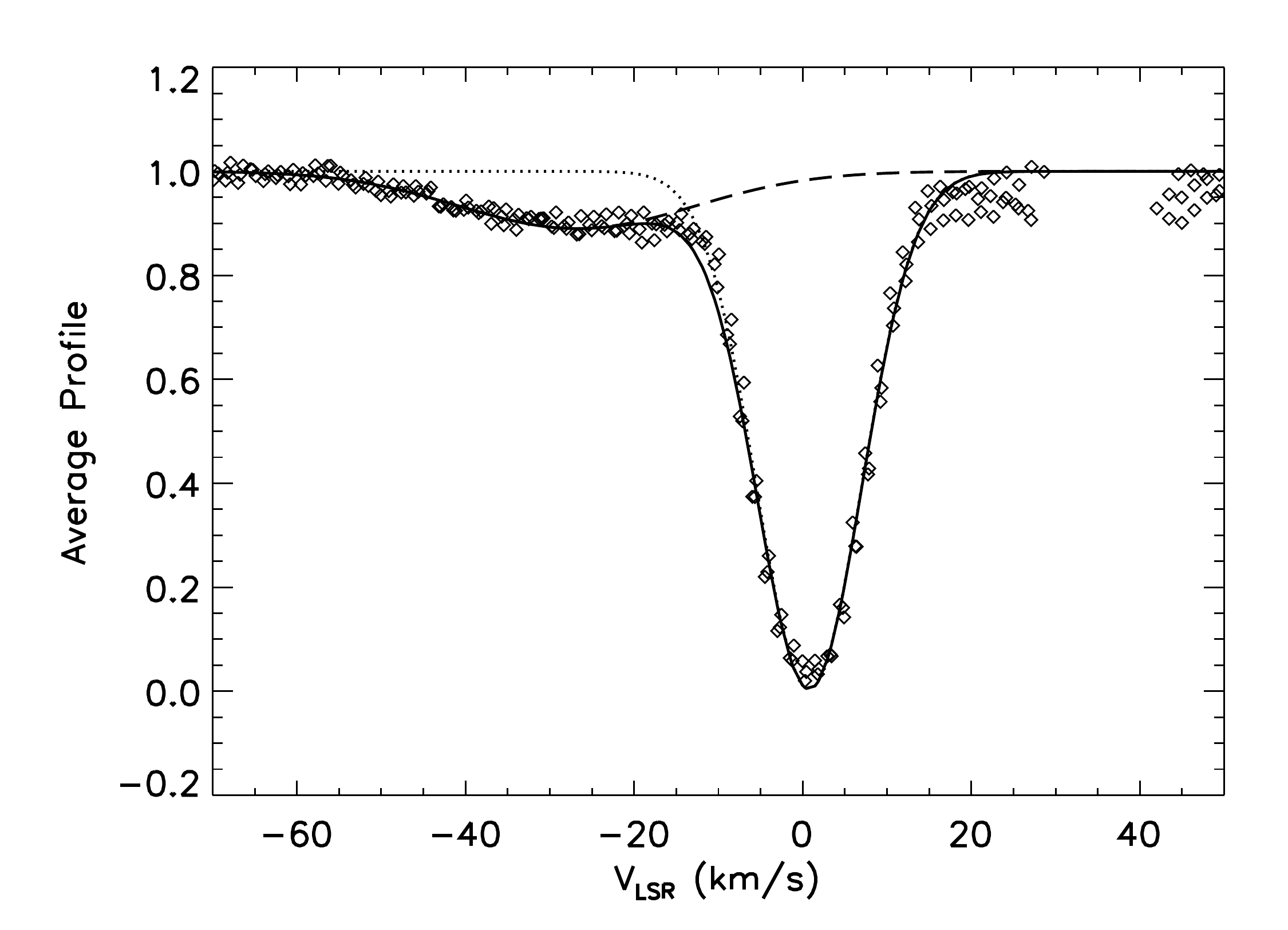}
\end{center}
\caption{$^{12}$CO line profile for the VV CrA IRC showing an optically thick,
low-J transitions (dotted line; v${}_{LSR}=0.78\mbox{ km s}^{-1}$,
FWHM=14.8$\mbox{ km s}^{-1}$) and a broad, optically thin line (dashed
line; v${}_{LSR}$=-24$\mbox{ km s}^{-1}$, FWHM=32.9$\mbox{ km s}^{-1}$).
The solid line is the combined fit.}

\end{figure}

\clearpage{}%

\begin{figure}
\includegraphics[angle=90,scale=0.35]{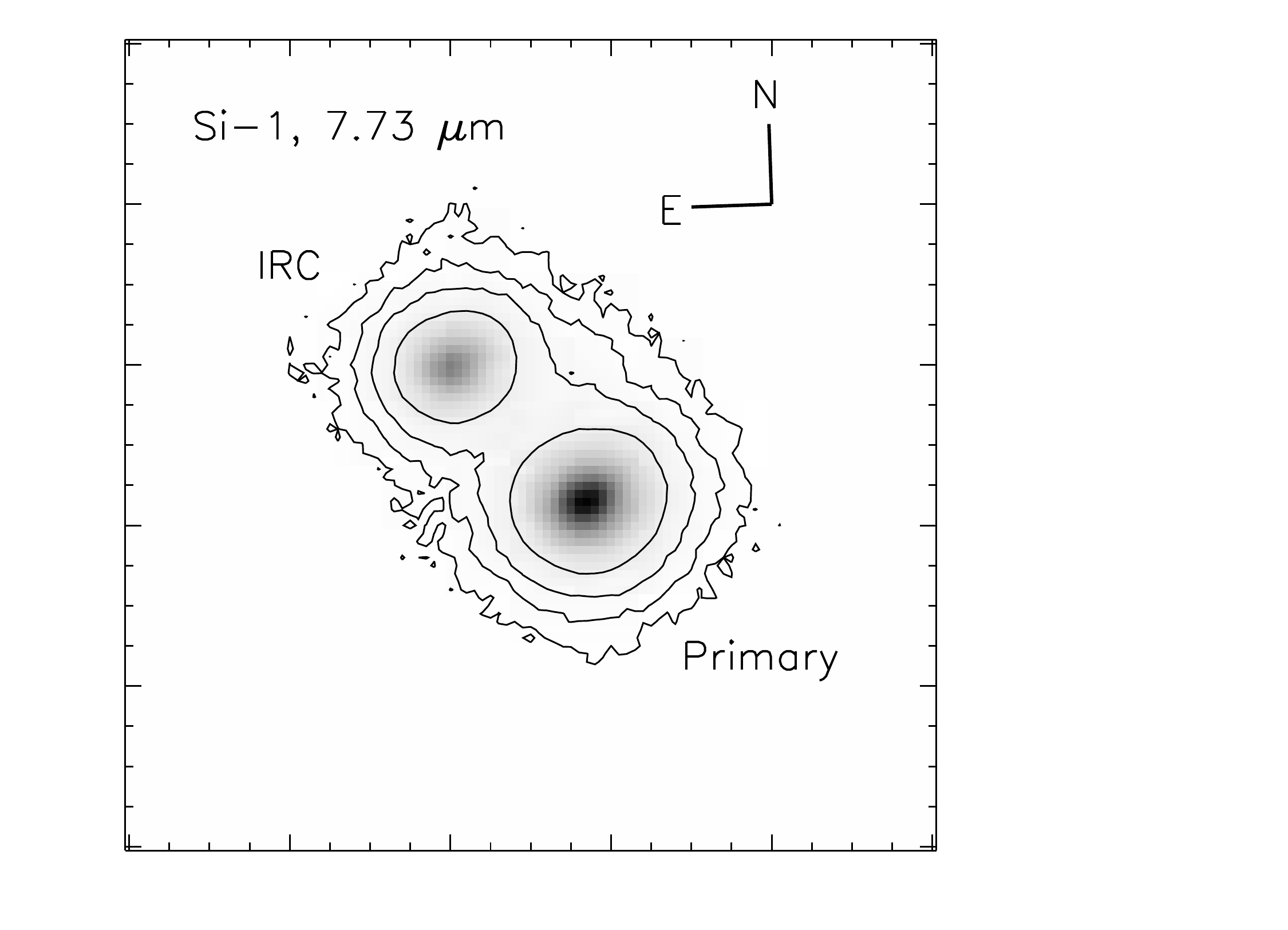}\includegraphics[angle=90,scale=0.35]{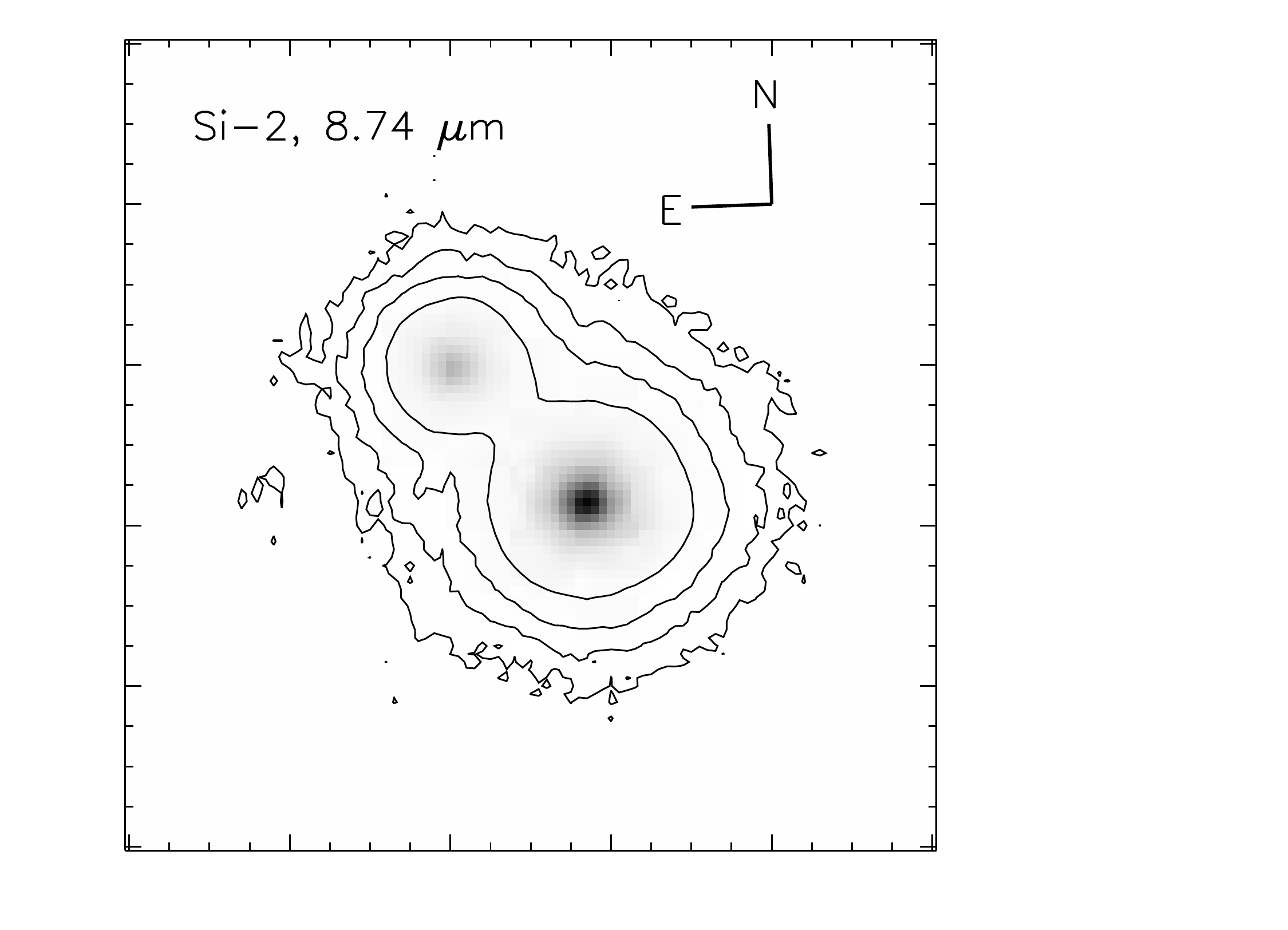}

\includegraphics[angle=90,scale=0.35]{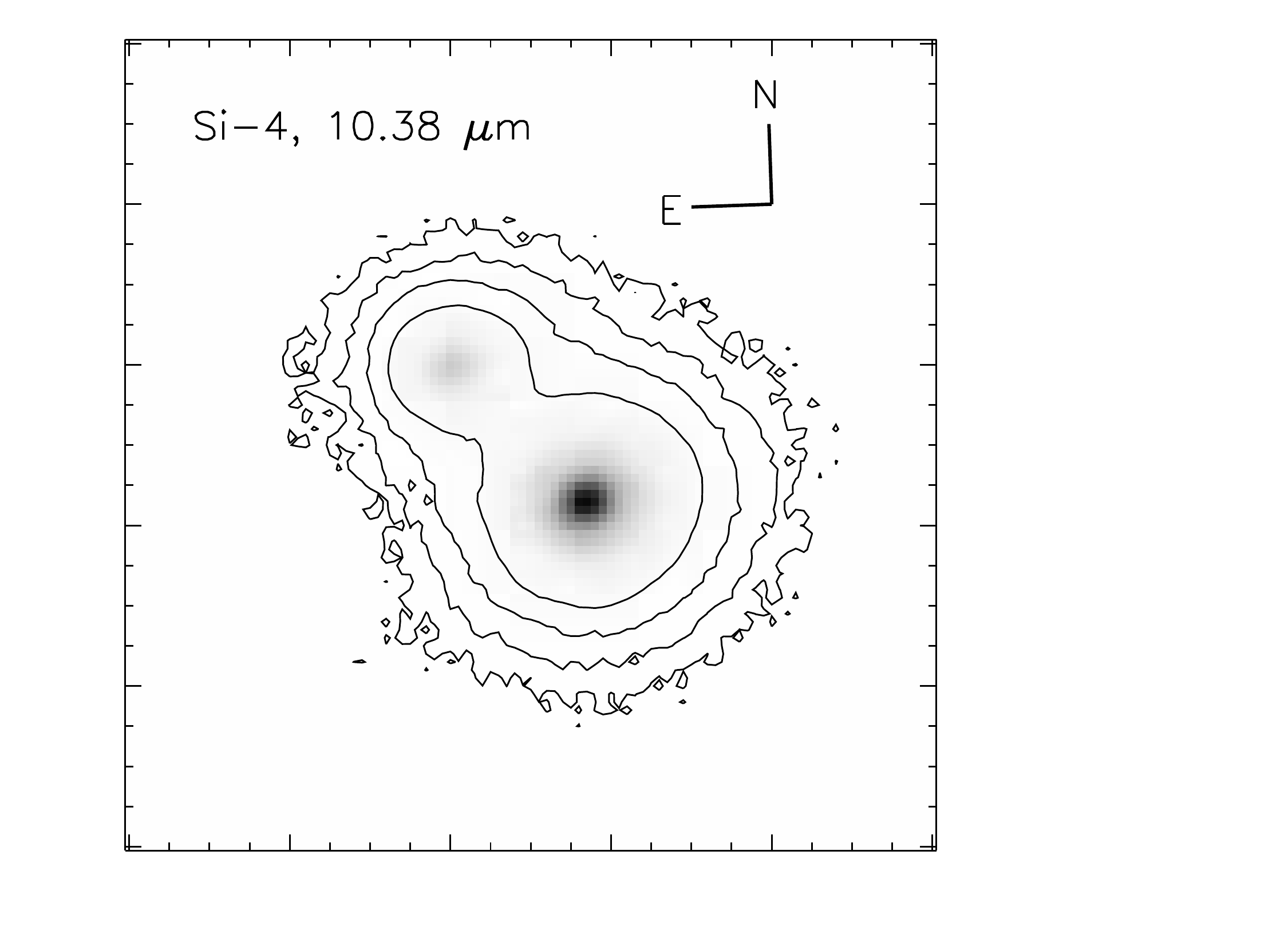}\includegraphics[angle=90,scale=0.35]{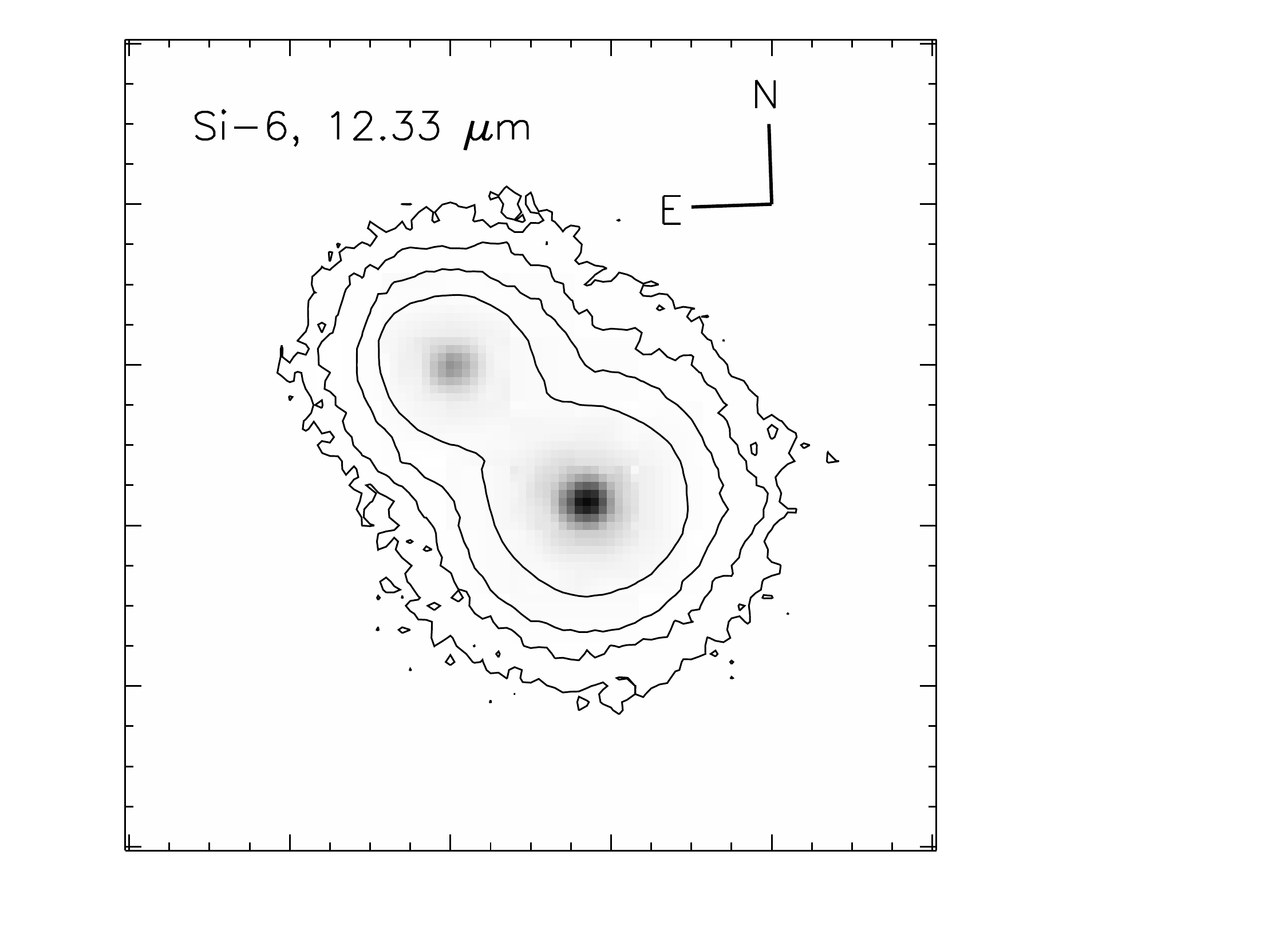}

\caption{Images of VV CrA taken with T-ReCS in the Si-1 (7.73 \textgreek{m}m),
Si-2 (8.74 \textgreek{m}m), Si-4 (10.38 \textgreek{m}m), and Si-6
(12.33 \textgreek{m}m) filters with the IRC being the NE component.
The contours are drawn at 5, 10, 20 and 50$\sigma$. The 5$\sigma$
contours, in Jy/pixel, correspond to: $5\sigma_{Si-1}$=0.080, $5\sigma_{Si-2}$=0.020,
$5\sigma_{Si-4}$=0.023, $5\sigma_{Si-6}$=0.038.}

\end{figure}

\clearpage{}%

\begin{figure}
\begin{center}
\includegraphics[angle=90,scale=0.7]{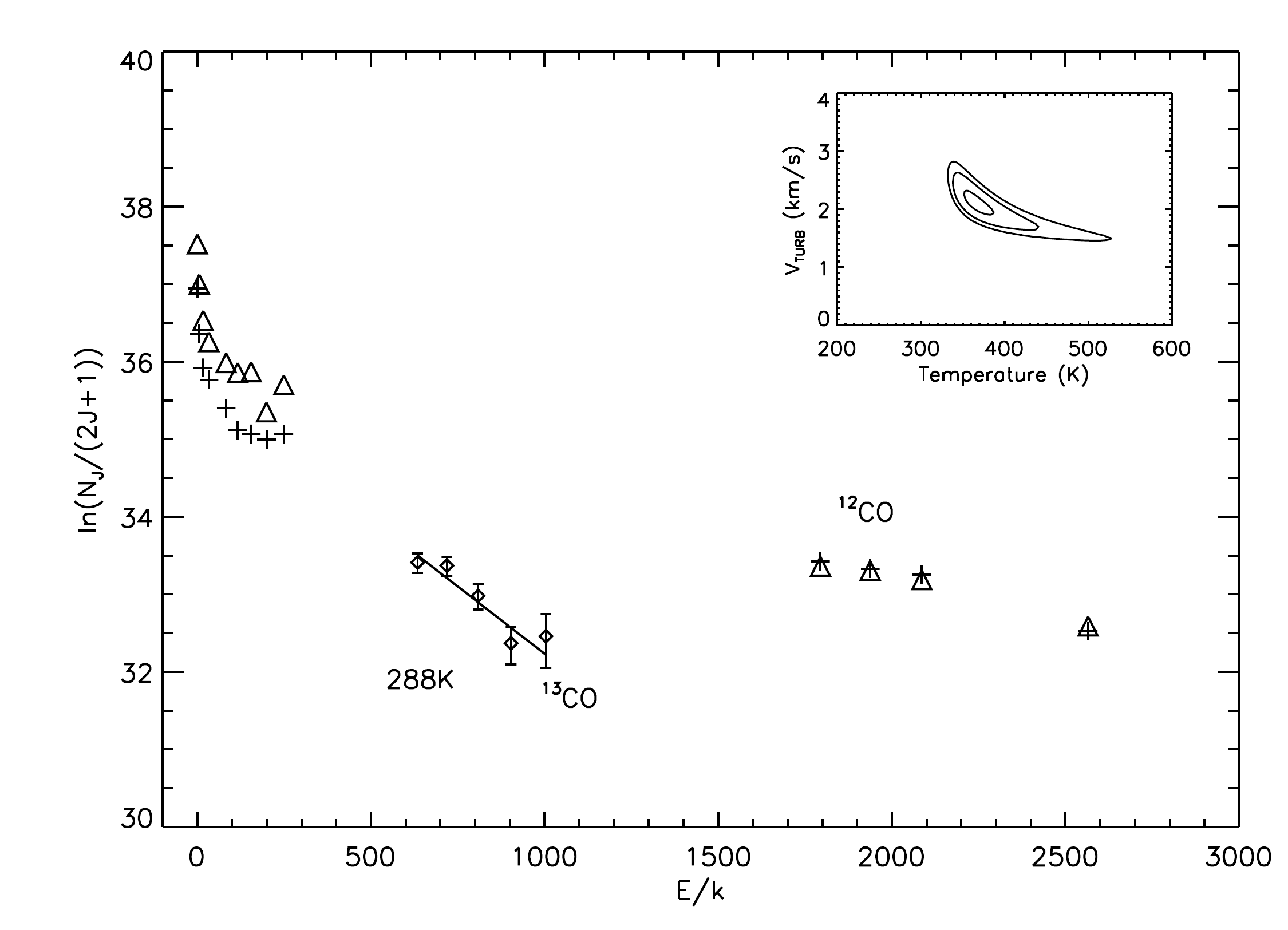}
\end{center}
\caption{Population diagram of $^{12}$CO and $^{13}$CO absorption as seen
toward DG Tau B. The $^{13}$CO (diamonds) is modeled as being optically
thin with temperature $\mbox{288 K}$. The $^{12}$CO line transitions
are optically thick (shown with triangles, indicating these are lower
limits) and we model the gas (crosses) with column density N($^{12}$CO)=$2.42\times10^{19}\mbox{ cm}^{-2}$,
T$=365\mbox{ K}$ and v$_{turb}=2.1\mbox{ km s}^{-1}$. $Inset:$
Goodness-of-fit contours for the temperature and turbulence velocity
of the $^{12}$CO P(25-27, 30) absorption lines seen toward DG Tau
B assuming a column density $2.42\times10^{19}\mbox{ cm}^{-2}$. The
contours show the 68, 95, and 99\% confidence levels. }

\end{figure}

\clearpage{}%

\begin{figure}
\begin{center}
\includegraphics[angle=90,scale=0.4]{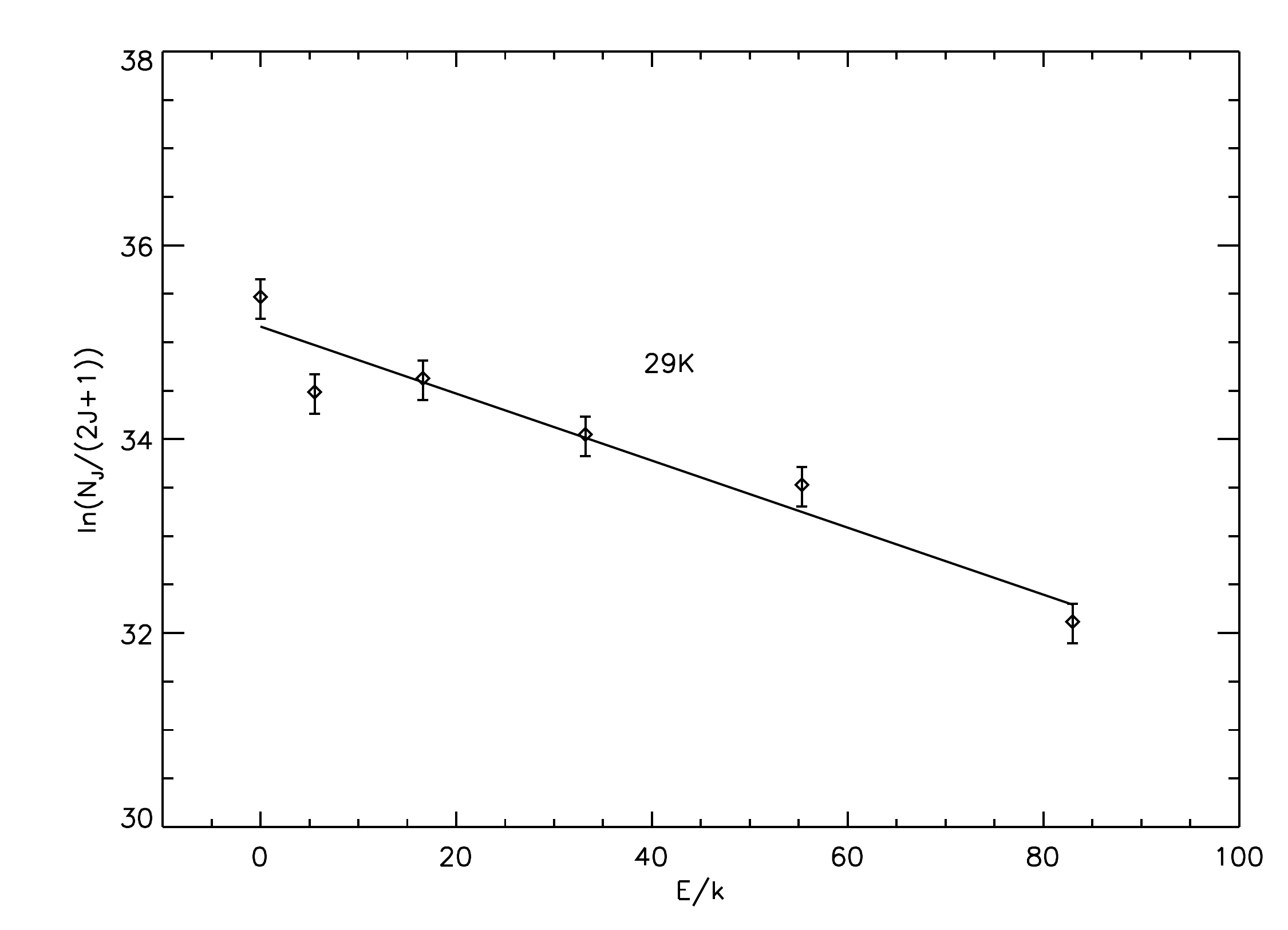}
\end{center}
\caption{Population diagram of $^{12}$CO absorption seen toward the VV CrA
Primary as measured using CRIRES spectra.}

\end{figure}

\clearpage{}%

\begin{figure}
\begin{center}
\includegraphics[angle=90,scale=0.4]{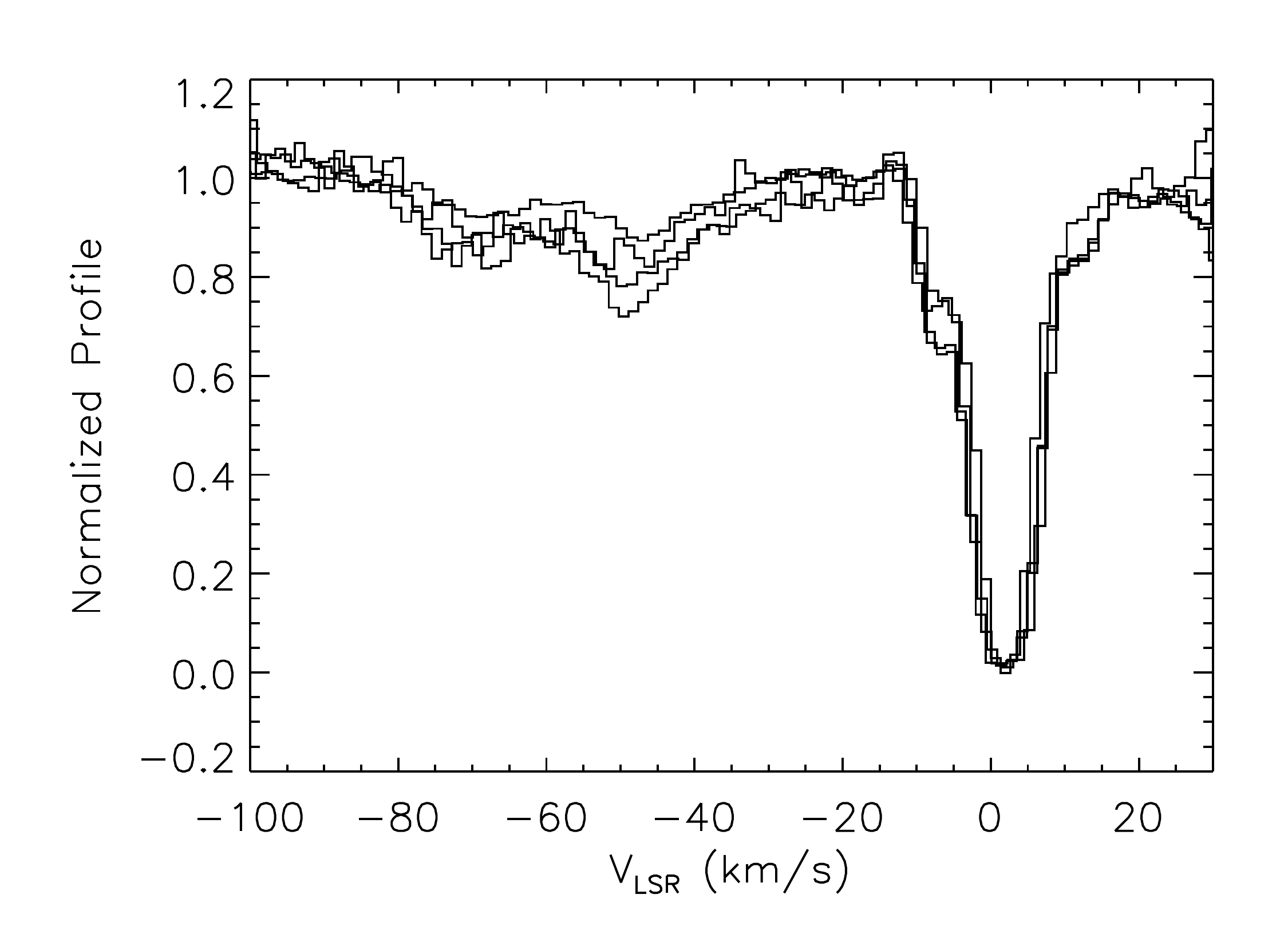}
\end{center}
\caption{$^{12}$CO line profile for the VV CrA IRC found with CRIRES, using
the P(3, 6, 7, 14) transitions as examples. The broad absorption found
with Phoenix is absent, replaced by other $^{12}$CO absorption lines.}

\end{figure}
\clearpage{}

\begin{figure}
\begin{center}
\includegraphics[angle=90,scale=0.4]{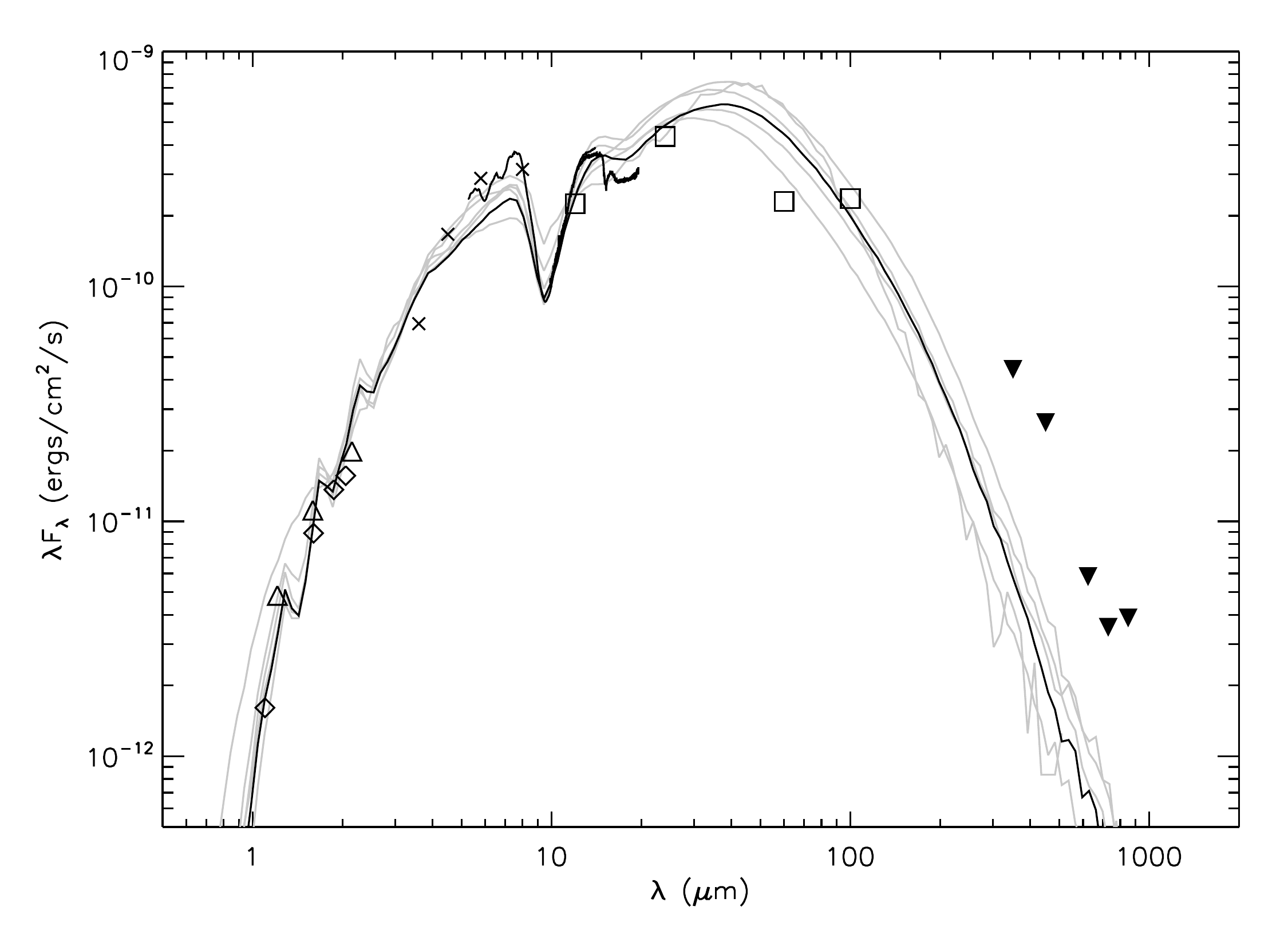}
\end{center}
\caption{Model fits to the SED of DG Tau B with best fit model shown in black
and subsequent fits shown in grey. Shown are data from Spitzer IRS
(continuum), HST (diamonds), 2MASS (triangles), IRAC (crosses; Luhman
et al. 2010), IRAS (squares; Weaver \& Jones 1992), and upper limits (upside-down arrows)
from SHARC (Andrews \& Williams 2005), SCUBA (Andrews \& Williams
2005), and CSO (Beckwith \& Sargent 1991).}

\end{figure}

\clearpage{}

\begin{sidewaysfigure}
\includegraphics[scale=0.6]{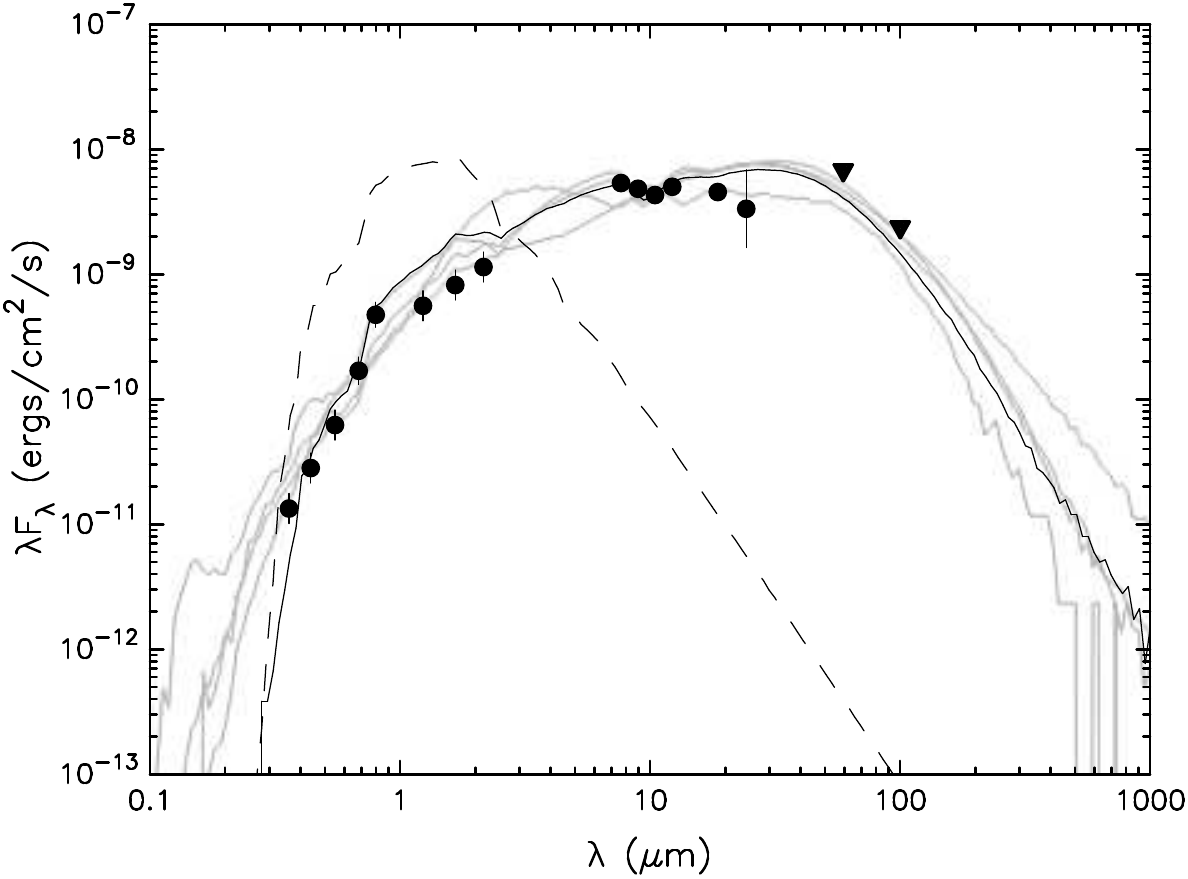}\includegraphics[scale=0.6]{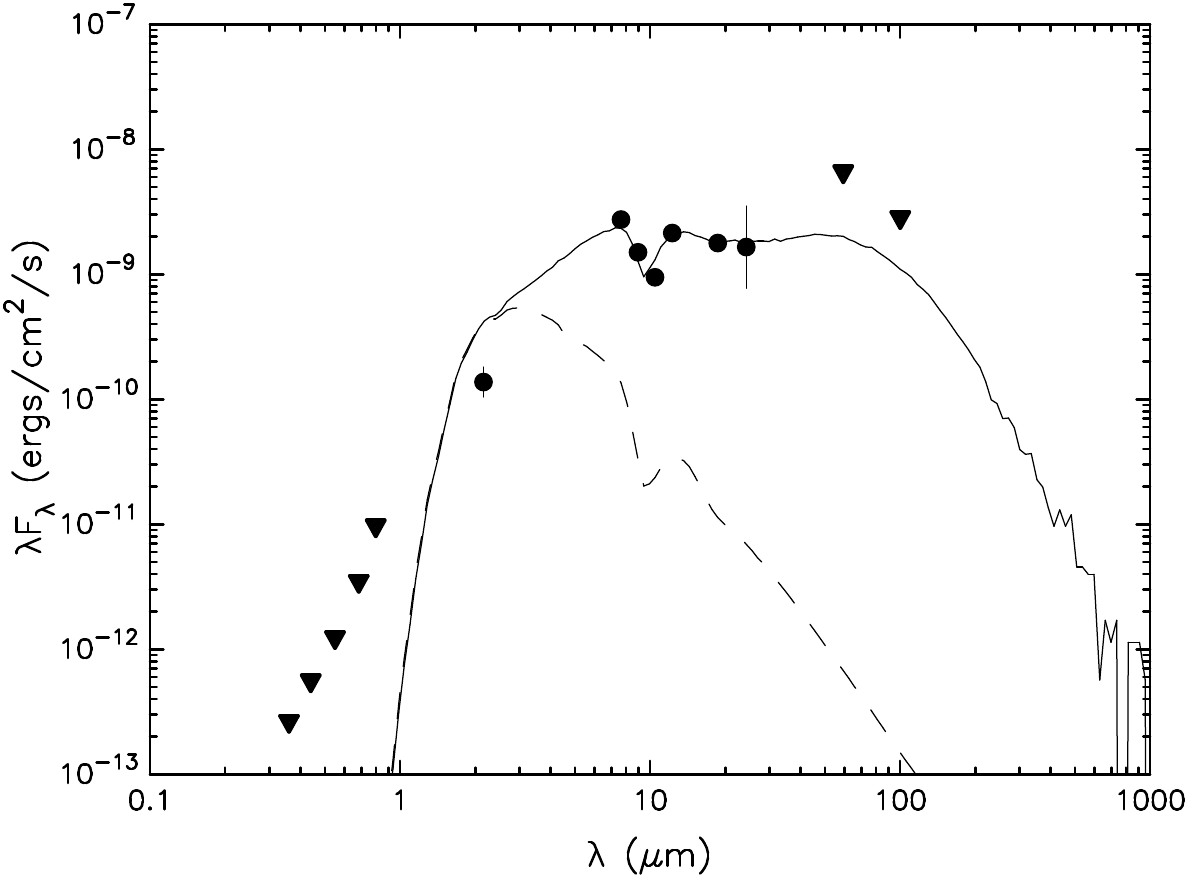}\includegraphics[scale=0.6]{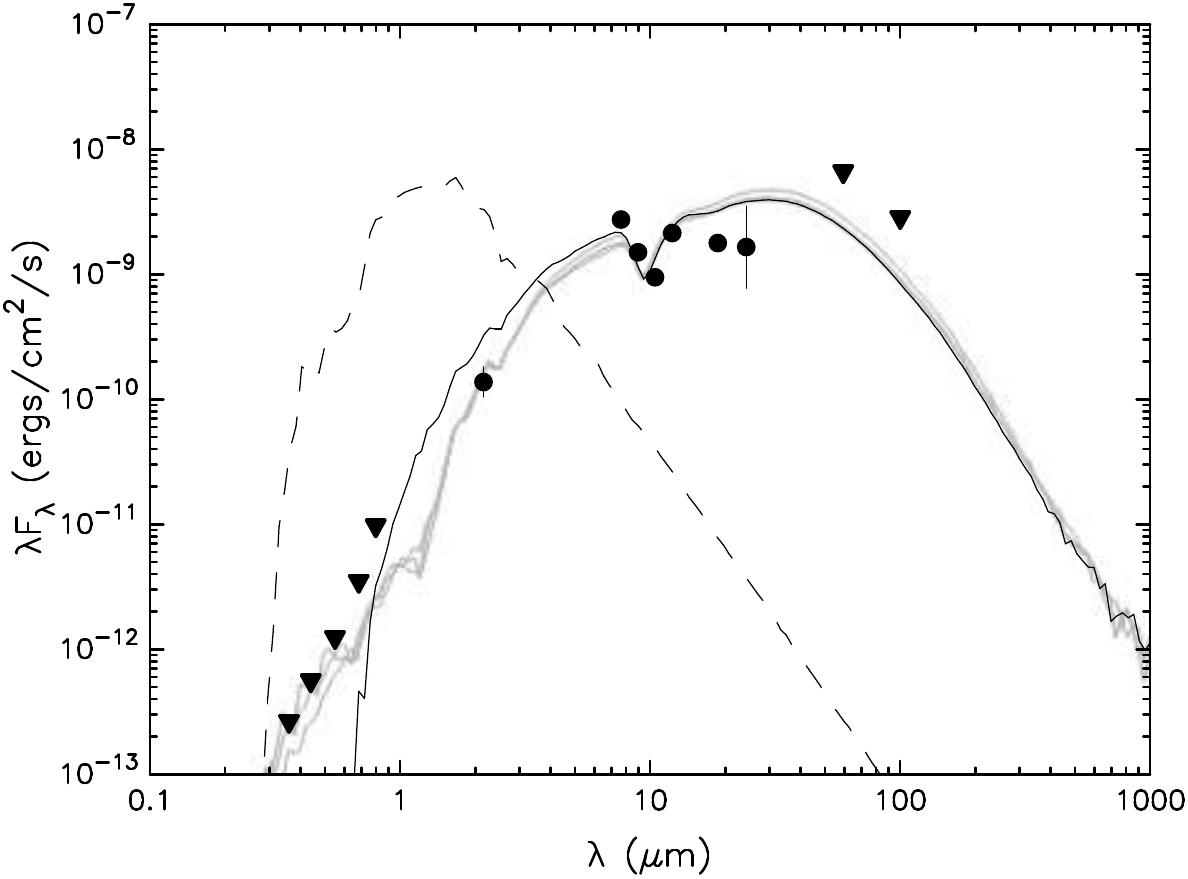}

\caption{Model fits to SEDs of the VV CrA Primary (left) and the VV CrA IRC
for two Cases: (middle; A) the Primary\textquoteright{}s disk is in
the line of sight to IRC; (right; B) the IRC disk is the main source
of extinction. See text for detailed explanation of spectral points (circles) and upper limits (triangles) used. The best fit model is shown in black and subsequent
fits in grey. The dashed line is the stellar photosphere emission
for the best fit model if viewed without circumstellar dust, but including
interstellar extinction.}

\end{sidewaysfigure}




\end{document}